\documentclass[aps,prl,twocolumn,groupedaddress,superscriptaddress]{revtex4-1}
\usepackage[english]{babel}
\usepackage{etoolbox}
\usepackage[capitalize]{cleveref}

\usepackage{graphicx,amsmath, amsfonts, amssymb, amsthm, mathrsfs, braket, bbm, bm, times,txfonts,color,verbatim}
\usepackage{hyperref}

\newcommand{\tr}{\mathrm{Tr}}

\newcommand{\di}{i}
\providecommand{\av}[1]{\langle #1 \rangle}
\providecommand{\sprod}[2]{\langle #1|#2 \rangle}
\providecommand{\ketbra}[2]{\ket{#1}\bra{#2}}

\begin{document}
\title{Towards superresolution surface metrology: Quantum estimation of angular and axial separations}

\author{Carmine Napoli}\email{carmine.napoli@nottingham.ac.uk}
\affiliation{School of Mathematical Sciences and Centre for the Mathematics and Theoretical Physics of Quantum Non-Equilibrium Systems, University of Nottingham, University Park Campus, Nottingham NG7 2RD, United Kingdom}
\affiliation{Manufacturing Metrology Team, Faculty of Engineering, University of Nottingham, Jubilee Campus, Nottingham NG8 1BB, United Kingdom}

\author{Samanta Piano}\email{samanta.piano@nottingham.ac.uk}
\affiliation{Manufacturing Metrology Team, Faculty of Engineering, University of Nottingham, Jubilee Campus, Nottingham NG8 1BB, United Kingdom}

\author{Richard Leach}\email{richard.leach@nottingham.ac.uk}
\affiliation{Manufacturing Metrology Team, Faculty of Engineering, University of Nottingham, Jubilee Campus, Nottingham NG8 1BB, United Kingdom}

\author{Gerardo Adesso}\email{gerardo.adesso@nottingham.ac.uk}
\affiliation{School of Mathematical Sciences and Centre for the Mathematics and Theoretical Physics of Quantum Non-Equilibrium Systems, University of Nottingham, University Park Campus, Nottingham NG7 2RD, United Kingdom}

\author{Tommaso Tufarelli}\email{tommaso.tufarelli@nottingham.ac.uk}
\affiliation{School of Mathematical Sciences and Centre for the Mathematics and Theoretical Physics of Quantum Non-Equilibrium Systems, University of Nottingham, University Park Campus, Nottingham NG7 2RD, United Kingdom}

\begin{abstract}
We investigate the localization of two incoherent point sources with arbitrary angular and axial separations in the paraxial approximation. By using quantum metrology techniques, we show that a simultaneous estimation of the two separations is achievable by a single quantum measurement, with a precision saturating the ultimate limit stemming from the quantum Cram\'er-Rao bound. Such a precision is not degraded in the sub-wavelength regime, thus overcoming the traditional limitations of classical direct imaging derived from Rayleigh's criterion. Our results are qualitatively independent of the point spread function of the imaging system, and quantitatively illustrated in detail for the Gaussian instance. This analysis may have relevant applications in three-dimensional surface measurements.
\end{abstract}

\date{\today}

\maketitle

{\bfseries \em Introduction.}---
High-resolution imaging is a cornerstone of modern science and engineering, which has enabled revolutionary advances in astronomy, manufacturing, biochemistry, and medical diagnostics. In traditional direct imaging based on classical wave optics, two incoherent point sources with angular separation smaller than the wavelength of the emitted light cannot be resolved due to fundamental diffraction effects \cite{rayleigh1879xxxi}, a phenomenon recently dubbed ``Rayleigh's curse'' \cite{tsang2016quantum}. Several techniques, including most prominently fluorescence microscopy \cite{fluorescencenobel}, have been introduced in recent years to overcome this limitation and achieve sub-wavelength imaging \cite{moerner2007new,leach2014applications}. Nevertheless, to determine the ultimate limits of optical resolution one needs to resort to a full quantum mechanical description of the imaging process \cite{genovese2016real}. In this respect, a breakthrough has been reported in a series of works \cite{tsang2016quantum,nair2016interferometric,nair2016far,lupo2016ultimate,tsang2017subdiffraction,kerviche2017fundamental,ang2017quantum,chrostowski2017super,rehavcek2017multiparameter,rehacek2017optimal,rehacek2017optimal2,zhou2018modern,tsang2019quantum} initiated by Tsang and collaborators \cite{tsang2016quantum}, who employed techniques from quantum metrology \cite{helstrom1976quantum,braunstein1994statistical,paris2009quantum,giovannetti2011advances} to prove that the achievable error in estimating the angular separation of two incoherent point sources, in the paraxial approximation, is in fact independent of said separation (no matter how small), provided an optimal detection scheme is performed on the image plane. These results, which stem from the fundamental quantum Cram\'er-Rao bound \cite{helstrom1976quantum,braunstein1994statistical} and {\em de facto} banish Rayleigh's curse \cite{tsang2016quantum}, have been  corroborated by proof-of-principle experiments \cite{yang2016far,paur2016achieving,tang2016fault,tham2017beating}.

The majority of the studies presented so far on quantum superlocalization, however, were limited to the case of point sources aligned on the same object plane, thus neglecting their  axial separation. The optical lateral resolution of an imaging system is an important characteristic, but it is not the only figure of merit relevant for the measurement of non-flat surfaces \cite{leach2015open}. When probing surface topography, the spacing of the points in an image must be considered, along with the ability to accurately determine the heights of features. In other words, the lateral resolution must be considered in conjunction with the ability of the system to transfer surface amplitudes \cite{de2012lateral}.

\begin{figure}[t!]
\includegraphics[width=8cm]{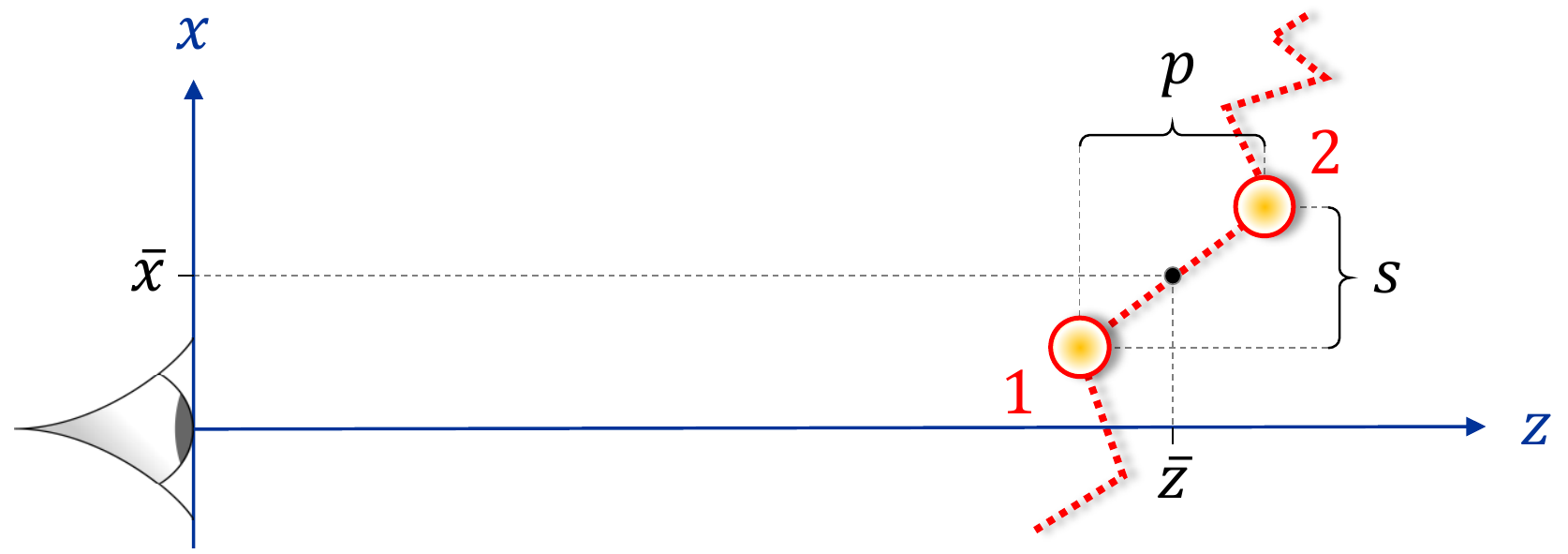}
\caption{Schematic of the two  sources. The four parameters to be estimated are: the angular separation $s$, the axial separation~$p$, the angular centroid coordinate $\bar{x}$, and  the axial centroid coordinate $\bar{z}$.\label{fig:system}}
\end{figure}

To address this issue, here we consider the simultaneous estimation of both angular and axial separations, as well as the corresponding centroid coordinates, of two incoherent point sources aligned in general on different object planes. These point sources may model, e.g., two emitters at the edges of a steep section on a rough surface, as indicated by the red dotted outline in \figurename~{\ref{fig:system}}.

We tackle the problem by resorting to the toolbox of {\em multiparameter quantum metrology}, a branch of quantum technology which is attracting increasing interest thanks to its prominent role in fundamental science and applications \cite{helstrom1976quantum,braunstein1994statistical,matsumoto2002new,paris2009quantum,giovannetti2011advances,toth2014quantum,szczykulska2016multi,ragy2016compatibility,braun2017quantum,pezze2016quantum,
monras2011measurement,genoni2013optimal,steinlechner2013quantum,pinel2013quantum,humphreys2013quantum,crowley2014tradeoff,vidrighin2014joint,banchi2015quantum,baumgratz2016quantum,altorio2016metrology,pirandola2017ultimate,pezze2017optimal,roccia2017entangling,yousefjani2017estimating,nichols2017multiparameter,proctor2018multiparameter,nair2018quantum}. We find that Rayleigh's curse does not occur even when the sources have a nonzero axial separation, and both axial and angular distances can be estimated simultaneously and with distance-independent precision by means of a single optimal quantum measurement, meeting the compatibility requirements for saturation of the multiparameter quantum Cram\'er-Rao bound \cite{matsumoto2002new,ragy2016compatibility}. These results are obtained analytically and are valid for any point spread function of the imaging system obeying the paraxial wave equation. We then specialize to the illustrative case of a Gaussian point spread function, and derive closed formulas for the achievable estimation error and its scaling with the parameters of interest as determined by the quantum Fisher information matrix,  showing that in the limit of small angular and axial distances all the parameters, including the centroid coordinates, become statistically independent.

{\bfseries \em Sources and imaging system model.}---
We approach the problem of estimating both axial and angular separation of two point sources by following a similar approach to Ref.~\cite{tsang2016quantum}, which is in turn inspired by Rayleigh's work \cite{rayleigh1879xxxi}. We assume that the detectable light on the image plane can be described as an incoherent mixture of two quasi-monochromatic scalar paraxial waves, one coming from each source. As shown in \figurename~{\ref{fig:system}}, our two sources are in general not lying on the same object plane (an `object plane' is a plane perpendicular to the optical axis $z$), and they feature an angular separation $s$ and an axial separation $p$.

Considering thermal sources at optical frequencies, we divide the total emission time into short coherence time intervals $\tau_c$, so that within each interval the sources can be assumed weak, i.e., effectively emitting at most one photon. This is a standard approach for modelling incoherent thermal sources \cite{labeyrie2006introduction, zmuidzinas2003cramer, goodman2015statistical, mandel1995optical, mandel1959fluctuations, gottesman2012longer, tsang2011quantum}, and it allows us to describe the quantum state $\rho$ of the optical field on the image plane as a mixture of a zero-photon state $\rho_0$ and a one-photon state $\rho_1$ in each time interval (neglecting contributions from higher photon numbers) \footnote{However, the assumption of weak sources can be lifted as in \cite{nair2016far,lupo2016ultimate}, leading to similar qualitative results for thermal sources of arbitrary intensity.},
\begin{equation}
\rho=\left(1-\varepsilon\right)\rho_0+\varepsilon\rho_1+o\big(\varepsilon^2\big),\label{rho}
\end{equation}
where $\varepsilon \ll 1$ is the average number of photons impinging on the image plane.
In practice, a detectable signal is obtained by measuring the optical field for a time $t\gg \tau_c$, so that many coherence time intervals are included, resulting in a non-negligible mean photon number.

We assume in general that the image-plane field amplitude generated by each source takes the form
\begin{equation}
\Psi_j\left(x,y\right)\equiv \psi(x-x_j,y,z_j),\label{tuttepsi}
\end{equation}
where $(x,y)$ are the image-plane coordinates, $(x_j,z_j)$ are the unknown coordinates of the sources $j=1,2$, $x_j$ being the coordinate perpendicular to the optical axis and $z_j$ the axial distance to the image plane (in this work we assume that the other coordinate $y_j=0$ is known). The amplitude function $\psi(x,y,z)$ obeys a paraxial wave equation of the form
\begin{equation}
i\partial_z\psi=G \psi,
\end{equation}
where $G$ is a self-adjoint differential operator featuring only $x$ and $y$ derivatives --- for example in free space one would have $G =\tfrac{1}{2k}(\partial_x^2+\partial_y^2)+k$, $k$ being the wavenumber. Since $[G,\partial_x]=0$, it follows $\Psi_j\left(x,y\right)=\exp(-iG z_j-x_j\partial_x)\psi(x,y,0)$.

We shall indicate with $a\left(x,y\right)$ the field annihilation operator at position $(x,y)$ on the image plane, satisfying the bosonic commutation rule $\left[a\left(x,y\right),a^\dagger\left(x',y'\right)\right]=\delta\left(x-x'\right)\delta\left(y-y'\right)$.

 We can then write the state $\rho_1$ as the incoherent mixture
 \begin{equation}\label{eqn:rho1}
 \rho_1 = \frac12 \left( \ket{\Psi_{1}}\bra{\Psi_{1}} + \ket{\Psi_2}\bra{\Psi_2}\right)\,,
 \end{equation}
 where the quantum state of the optical field on the image plane corresponding to the emission of one photon by the source $r$ may be expressed as
\begin{align}
\ket{\Psi_j}&=\exp\left(-iG z_j-x_j\partial_x\right)\ket{\psi},\label{Ztrasl}\\
\ket{\psi}&\equiv\int_{\mathbb{R}^2}\psi\left(x,y,0\right)a^\dagger\left(x,y\right)\ket{0}\,dx\,dy\,,
\end{align}
$\ket{0}$ being the field vacuum state. Finally, we may take $\psi(x,y,0)$ real, which results in some simplifications later on. This can be assumed without loss of generality, as the complex phase of $\psi(x,y,0)$ may be compensated by a redefinition of $a(x,y)$ that is independent of the source parameters. However, $\psi(x,y,z)$ will have in general a nontrivial phase profile.

{\bfseries \em Multiparameter estimation and quantum Cram\'er-Rao bound.}---
We work under the assumption that the photon statistics of our sources is Poissonian, following a similar approach as in Ref.~\cite{tsang2016quantum}. We can thus assume that in a single run of the experiment, which lasts for $M$ coherence time intervals, $M$ copies of the state $\rho$ in Eq.~\eqref{rho} are prepared and measured (equivalently, one may consider the input state $\rho^{\otimes M}$). On average, this yields $M\varepsilon$ photons per run. In order to apply the standard tools of estimation theory, we further assume that $\nu\gg1$ runs are performed, after which the measurement data are processed to build estimators for the unknown parameters.

In our case, the parameters of interest are the angular and axial relative coordinates and the centroid coordinates of the sources, indicated as $s,\bar{x},p,\bar{z}$, see \figurename~{\ref{fig:system}}. 
We thus write the state $\rho$ as a function of four parameters $\{\lambda_\mu\}_{\mu=1,\ldots,4}$, where
\begin{equation}\label{eqn:parameters}
\begin{split}
\lambda_1&\equiv s=x_2-x_1,\quad\lambda_2\equiv \bar{x}=\frac{x_2+x_1}2,\\
\lambda_3&\equiv p=z_2-z_1,\quad\lambda_4\equiv \bar{z}=\frac{z_2+z_1}2,\\
\end{split}
\end{equation}

The statistical error (variance) $\Delta\lambda_\mu^2$ of any {\em unbiased} estimator of the unknown parameter $\lambda_\mu$ is lower bounded via the quantum Cram\'er-Rao bound (qCRb) \cite{helstrom1976quantum,braunstein1994statistical}
\begin{equation}
\sum_{\mu=1}^{4}\Delta\lambda_\mu^2\geq\frac{1}{\nu M \varepsilon}\tr[H^{-1}],
\label{QCB_ottimista}
\end{equation}
where $H$ is the quantum Fisher information matrix (qFim) of the single-photon state $\rho_1$ (equivalently, this can be seen as the qFim per coherence time interval per photon). The prefactor on the RHS of Eq.~\eqref{QCB_ottimista} is obtained by exploiting the additivity property $\text{qFim}(\rho^{\otimes M})=M\times\text{qFim}(\rho)$, and by approximating that $\text{qFim}(\rho)\simeq\varepsilon \times\text{qFim}(\rho_1)$ at leading order in $\varepsilon$ (since the field vacuum state $\rho_0$ is independent of all source parameters and is always orthogonal to $\rho_1$ --- see also the discussion in the Appendix of Ref.~\cite{tsang2016quantum}). The resulting linear dependence on the total photon number $\nu M \varepsilon$ is characteristic of classical light sources \cite{giovannetti2011advances,tsang2015quantum}.

The qCRb suggests that, the higher the qFIm element $H_{\mu \mu}$, the more precisely (i.e., with lower statistical error) one may be able to estimate the parameter $\lambda_\mu$, by performing a suitable measurement. While for a single parameter the qCRb can always be saturated asymptotically by means of an adaptive procedure \cite{paris2009quantum}, this is no longer the case for multiparameter estimation, as the parameters may not always be  compatible \cite{ragy2016compatibility}; we will discuss this issue in detail later in the manuscript.

{\bfseries \em Results.}---
We recall that the qFim elements are given by
\begin{equation}\label{eqn:qfim}
H_{\mu\nu}={\rm Re}\left[\tr(\rho_1 L_\mu L_\nu)\right],
\end{equation}
where $L_\mu$ is the symmetric logarithmic derivative (SLD) for the parameter $\lambda_\mu$, defined implicitly by the equation
\begin{equation}\label{eqn:sldef}
2\frac{\partial \rho_1}{\partial \lambda_\mu}=L_\mu\rho_1+\rho_1 L_\mu.
\end{equation}
The following matrix (proportional to the averaged SLD commutators) will also be of interest for our  discussion,
\begin{equation}\label{eqn:comms}
\Gamma_{\mu\nu}\equiv{\rm Im}\left[\tr(\rho_1 L_\mu L_\nu)\right].
\end{equation}

For the problem under investigation, we have derived general analytical expressions for both matrices $H$ and $\Gamma$, as presented in detail in Appendix~A. Our derivation relies on the expansion of $\rho_1$ in the generally \textit{non-orthogonal} basis \begin{equation}\left\{\ket{\Psi_{1}},\ket{\Psi_{2}},\partial_{x_1}\ket{\Psi_{1}},\partial_{z_1}\ket{\Psi_{1}},\partial_{x_2}\ket{\Psi_{2}},\partial_{z_2}\ket{\Psi_{2}}\right\},\end{equation}
followed by standard linear algebraic manipulations. This method results in significant simplifications over previous studies of quantum superlocalization (typically relying on the explicit construction of an orthogonal basis to span the support of $\rho_1$ and its derivatives, as e.g.~in \cite{tsang2016quantum}), and may be of independent interest in its own right for the field of multiparameter quantum metrology. Thanks to the representation of $\ket{\Psi_j}$ given in Eq.~\eqref{Ztrasl}, it is easy to check that all the scalar products between the above basis vectors only depend on $s=x_2-x_1$ and $p=z_2-z_1$, which in turn implies that the qFim is independent of the centroid coordinates $\bar{x}$ and $\bar{z}$. The corresponding physical interpretation is that the information content of the emitted light is not affected by propagation along the optical axis, or by a redefinition of the image plane origin. Additional simplifications follow from our assumption $\psi(x,y,0)\in\mathbb{R}$, which implies $\braket{\psi|\partial_x\psi}=0$. We then find that the qFim is composed of the diagonal elements
\begin{align}\label{Q1}
&H_{ss}=\braket{\partial_x\psi|\partial_x\psi},\quad H_{pp}=\Delta G^2,\\
&H_{\bar{x}\bar{x}}=4\braket{\partial_x\psi|\partial_x\psi}-4(\partial_s |\gamma|)^2-4\frac{|\gamma|^2(\partial_s \varphi)^2}{1-|\gamma|^2},\\
&\begin{aligned}H_{\bar{z}\bar{z}}=\frac{4}{1-|\gamma|^2}\bigg\{&\Delta G^2-(\partial_p |\gamma|)^2-|\gamma|^2\Big[\av{G^2}-(\partial_p |\gamma|)^2\\
&+2\av{G}\,\partial_p \varphi+(\partial_p\varphi)^2\Big]\bigg\},
\end{aligned}\label{Q1end}
\end{align}
while the off-diagonal elements are all zero except
\begin{align}\label{Q2}
&H_{\bar{x}\bar{z}}=-\frac{4 |\gamma|^2 \left(\partial_s \varphi\right) \left(\av{G}+\partial_p \varphi\right)}{1-|\gamma|^2}-4 (\partial_s|\gamma|)(\partial_p|\gamma|).
\end{align}
At the same time, the only nonzero matrix elements of $\Gamma$ are
\begin{align}
\Gamma_{s\bar{x}}&=-\frac{2 |\gamma|^3 (\partial_s |\gamma|) (\partial_s \varphi)}{1-|\gamma|^2},\label{GG1} \\
\Gamma_{p\bar{z}}&=-\frac{2   |\gamma|^3  (\partial_p |\gamma|) \left(\av{G}+\partial_p \varphi)\right)}{1-|\gamma|^2}, \label{GG2}\\
\Gamma_{s\bar{z}}&=2  |\gamma| \left((\partial_p|\gamma|) (\partial_s \varphi)-\frac{\partial_s|\gamma| \left(\partial_s \varphi+\av{G}\right)}{1-|\gamma|^2}\right),\label{GG3}\\
\Gamma_{\bar{x}p}&=2  |\gamma| \left(-(\partial_s |\gamma|) \left(\av{G}+\partial_p \varphi\right)+\frac{(\partial_p |\gamma|)(\partial_s \varphi)}{1-|\gamma|^2}\right). \label{GG4}
\end{align}
The following shorthands have been used:
\begin{align}
\gamma&\equiv\braket{\Psi_1|\Psi_2},\quad \varphi\equiv{\text {arg}}\,\gamma,\\
\av{O}&\equiv\braket{\psi|O|\psi},\quad\Delta G^2\equiv\braket{G^2}-\braket{G}^2,
\end{align}
where we emphasize that $\gamma=\gamma(s,p)$ is the only quantity depending on the source coordinates.
A fundamental result can be immediately inferred from Eqs.~\eqref{Q1} and below: for any point spread function that satisfies the paraxial wave equation, $H_{ss}$ and $H_{pp}$ are {\em constant}. This statement exemplifies how our results provide new insights on the problem of sub-wavelength imaging, while correctly reproducing what is known for $p=0$ \cite{tsang2016quantum}. We note in particular that Rayleigh's curse {\em does not} affect the estimation of the angular separation $s$ nor that of the axial separation $p$.

Taking one step further, we can now investigate how close one can get to the limits imposed by the qCRb in practical experiments. In quantum estimation theory, multiparameter problems embody a nontrivial generalization of the single-parameter case \cite{matsumoto2002new,paris2009quantum,szczykulska2016multi, ragy2016compatibility}: if an estimation scheme is optimized for a particular parameter, it typically results into an increased error in estimating the others. However, in the best case scenario, such a trade-off does not apply, and one can identify an optimal protocol for the estimation of all the parameters simultaneously. This happens if and only if the parameters are {\em compatible}, i.e., they satisfy the following conditions \cite{ragy2016compatibility}: (i) There is a single probe state yielding the maximal qFim element for each of the parameters; (ii) There is a single measurement which is jointly optimal for extracting information on all the parameters from the output state, ensuring the asymptotic saturability of the qCRb; (iii) The parameters are statistically independent, meaning that the indeterminacy of one of them does not affect the error on estimating the others. We recall also that (ii) holds iff $\Gamma_{\mu,\nu}=0\quad\forall\mu\ne\nu$, while (iii) is equivalent to the condition $H_{\mu\nu}=0\quad\forall\mu\ne\nu$.

In this paper we do not focus on the first condition, since our theory is built around a realistic imaging scenario in which the emission properties of the sources are fixed in advance. Yet, it is worth investigating conditions (ii) and (iii), since they have crucial implications for the actual achievability of the statistical errors given by the qCRb.
Remarkably, we find that conditions (ii) and (iii) are {\em always satisfied} for the pair of parameters $(s,p)$ --- independently of the specifics of the point spread function. In the simplified scenario where $(\bar{x},\bar{z})$ are estimated independently or known in advance, it is thus possible to construct a physical measurement and estimation strategy for $s$ and $p$ saturating Eq.~\eqref{QCB_ottimista} asymptotically \cite{matsumoto2002new,ragy2016compatibility}. On the other hand, we can see that conditions (ii) and (iii) do not hold in general for the full set of parameters $(s,p,\bar{x},\bar{z})$. Yet, we shall see in the example below that there is at least one relevant type of point spread function for which conditions (ii) and (iii) are satisfied for all parameters in the limit $s\to0,p\to0$.

\begin{figure}[t!]
\centering\vspace*{-.3cm}
\includegraphics[width=7cm]{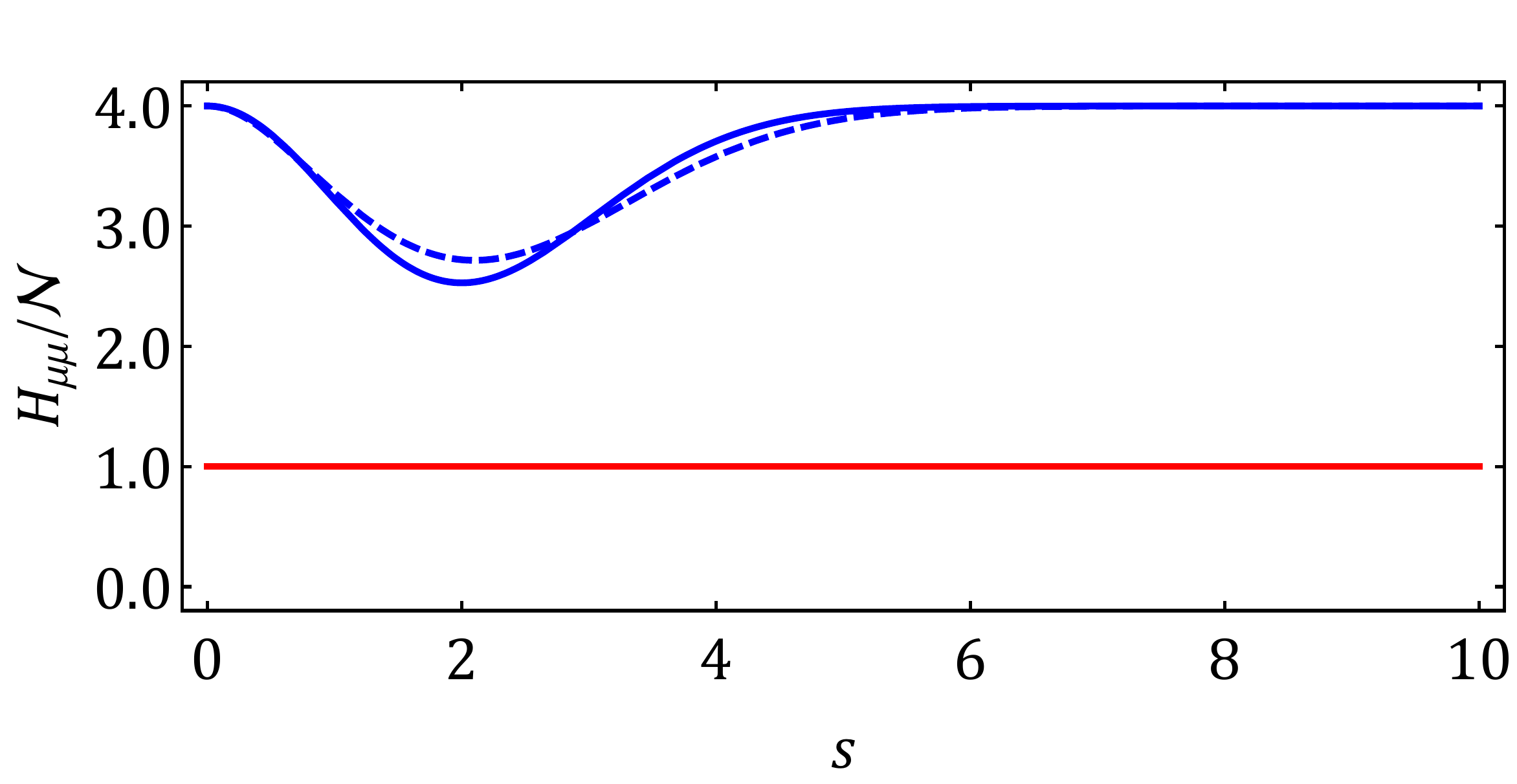}\\
\vspace*{-.1cm}\includegraphics[width=7cm]{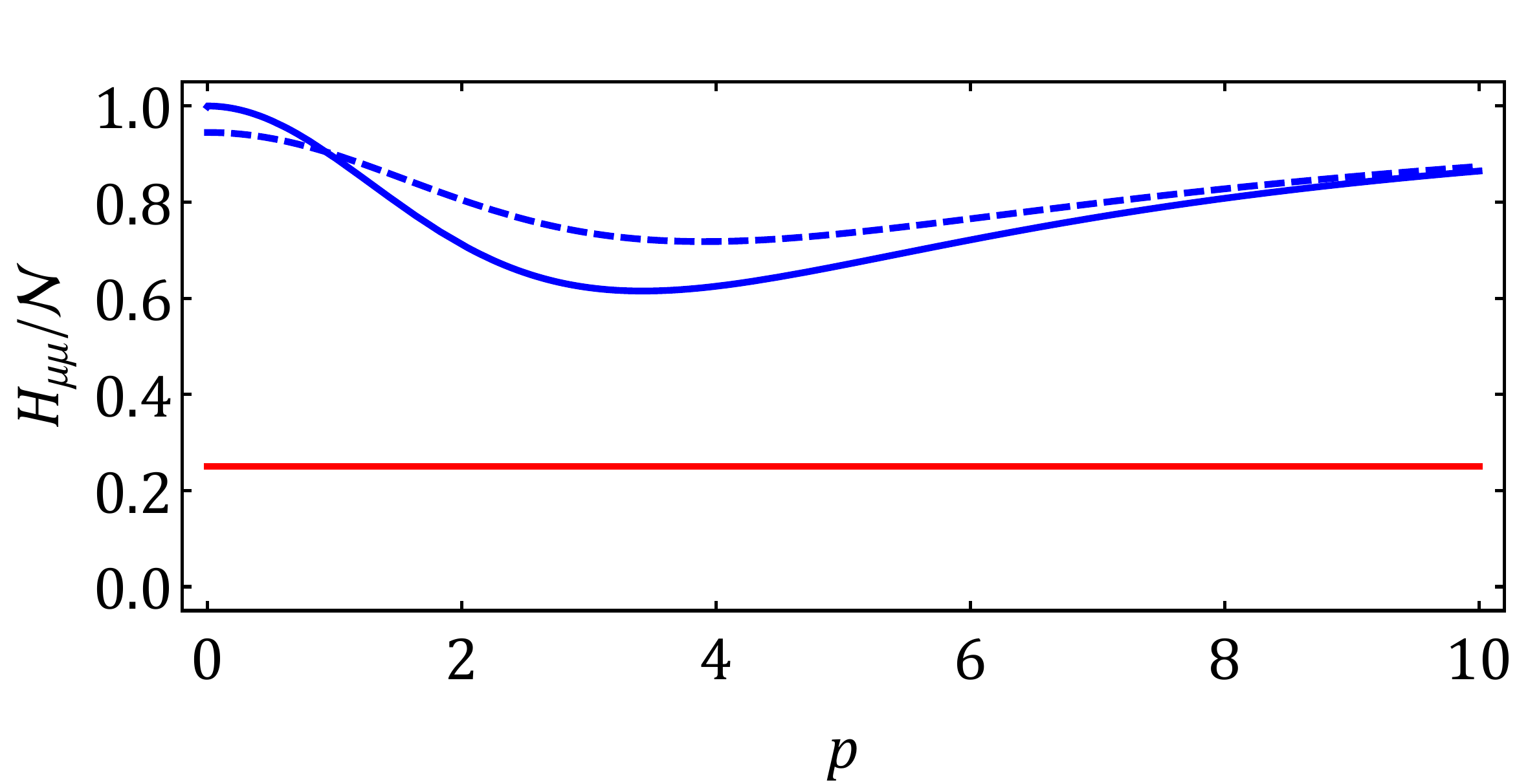}
\caption{{\bfseries \em Top: Angular localization.} Plots of the qFim elements  $H_{ss}$ (red, lowermost curve) and $H_{\bar{x}\bar{x}}$ (blue, uppermost curves), versus the angular separation $s$;  continuous lines refer to $p=0$ and dashed lines to $p=2$. {\bfseries \em Bottom: Axial localization.}
Plots of the qFim elements  $H_{pp}$ (red, lowermost curve) and $H_{\bar{z}\bar{z}}$ (blue, uppermost curves), versus the axial separation $p$; continuous lines are for $s=0$ and dashed lines for $s=1$. The results are for Gaussian beams with $k=1, z_R=2$. The vertical axes are normalized to ${\cal N} = \frac12 k/z_R$. 
}
\label{fig:mevsmankei}
\end{figure}

We consider in what follows a Gaussian beam in free space,
\begin{equation}
\psi(x,y,z)=\sqrt{\frac{k\, z_R}{\pi }}\frac{i}{z+iz_R} \exp \left(\frac{-ik \left(x^2+y^2\right)}{2 (z+iz_R)}-i k z\right),\label{gaussicspread}
\end{equation}
where $z_R$ is a length parameter characterizing the beam, typically assumed of the same order as the wavelength, i.e. $\sim1/k$.
Eq.~\eqref{gaussicspread} can be obtained, for example, if the fields generated by the two sources are well approximated by Gaussian beams in the vicinity of the image plane \cite{svelto1998principles}. We thus obtain
\begin{align}
&\gamma=\frac{2 i z_R}{p+2 i z_R} \exp\left(-ikp-\frac{i}{2}\frac{ks^2}{p+2 i z_R}\right), \quad
\braket{\partial_{x}\psi|\partial_{x}\psi}=\frac{k}{2z_R},\nonumber \\
&\braket{G}=k-\frac{1}{2z_R}, \quad
\braket{G^2}=k^2-\frac{k}{z_R}+\frac{1}{2 z_R^2}.
\end{align}
By substituting the above expressions in the qFim elements calculated previously, we find fully analytical closed formulas (as reported in Appendix~B) that allow us to perform a comprehensive analysis of the multiparameter estimation problem under investigation.
Furthermore, the Gaussian case bears the advantage that it can be easily compared with the existing literature that tackled the estimation of $s$ alone (typically fixing $p=0$). To support the solidity of our results, we have indeed checked that, in the limit $p\to0$, our expressions for $H_{ss}$ and $H_{\bar{x}\bar{x}}$ match the appropriate quantities in Refs.~\cite{tsang2015quantum,tsang2016quantum}.

Our results become particularly interesting in the regime $ks, kp \ll 1$, which is precisely the one of relevance to sub-wavelength imaging.
In the limit we have
\begin{align}\label{eqn:qfidiag}
\lim_{(s,p)\to(0,0)}H&=\mathrm{diag}\left\{\frac{k}{2z_R},\,\frac{2k}{z_R},\,\frac{1}{4z_R^2},\,\frac{1}{z_R^2}\right\},\\
\lim_{(s,p)\to(0,0)}\Gamma&= \mathrm{diag}\{0,0,0,0\}\,,\label{limiteGamma}
\end{align}
meaning that the four parameters $s,\bar{x},p,\bar{z}$  are approximately
statistically independent when the two sources have infinitesimal angular and axial separation.
The behaviour of the four diagonal qFim elements $H_{\mu \mu}$ as a function of the separations $s$ and $p$ is illustrated in Fig.~\ref{fig:mevsmankei}; the top panel can be compared directly with Fig.~2 of \cite{tsang2016quantum}. From the plots and from Eq.~\eqref{eqn:qfidiag}, we see that the qFIm diagonal elements tend to a nonzero value when $s,p \to 0$. Hence the fundamental lower bound on the total estimation error, $\propto \tr[H^{-1}]$, stays finite even when the two sources are infinitesimally close, instead of diverging as in direct imaging \cite{rayleigh1879xxxi,tsang2016quantum}. Eq.~\eqref{limiteGamma} further suggests that it should be possible to construct a single measurement that is approximately optimal for the estimation of \textit{all four parameters} when $ks,kp\ll1$. The construction of such a measurement will be addressed in future work.

{\bfseries \em Conclusions.}---
We determined the ultimate quantum limits to the simultaneous estimation of both angular and axial separations and centroid coordinates of two  incoherent point sources on different object planes in the paraxial approximation. Our results indicate that there exists a jointly optimal detection scheme that enables resolving the sources even when arbitrarily close, reasserting that Rayleigh's curse is merely an artefact of classical detection in direct imaging. In practice, a measurement apparatus approaching the optimal precision can be designed by adapting the methods of \cite{pezze2017optimal,rehacek2017optimal,rehacek2017optimal2,roccia2017entangling,yang2018optimal}, in particular extending the ``spatial-mode demultiplexing'' or ``superlocalization by image inversion interferometry'' techniques \cite{tsang2016quantum,nair2016interferometric} to the axially separated setting considered here.

While some of our findings were illustrated explicitly for Gaussian beams, our framework is general and can be applied to any point spread function that satisfied the paraxial wave equation, thanks to the exact expressions in Eqs.~(\ref{Q1})--(\ref{GG4}). This leads to qualitatively similar results as those presented here. In particular, the two most important conclusions, namely that the qFim elements for the angular distance $s$ and for the axial distance $p$ are both independent of $s$ and $p$, and that the joint estimation of $s$ and $p$ fulfils the measurement compatibility condition leading to the saturation of the quantum Cram\'er-Rao bound in Eq.~\eqref{QCB_ottimista}, are in fact valid for any point spread function.

\begin{acknowledgments}

This work constitutes an important application of multiparameter quantum estimation theory to a realistic imaging setting, extending the seminal work of Ref.~\cite{tsang2016quantum}. Our analysis, combined with the one in \cite{ang2017quantum}, yields a quantum enhanced toolbox for full 3D sub-wavelength localization. This paves the way to further experimental demonstrations and innovative metrology solutions in scientific, industrial and biomedical domains, such as sub-nanometre depth mapping in rough surfaces, and dynamical interaction analysis of heterogeneous molecules in a cellular environment \cite{moerner2007new,leach2014applications,leach2015open,wolf2011coherent}.

{\bfseries \em Note added.}---
Shortly after the initial submission of this work, quantum superresolution of two incoherent sources in three dimensions has been studied independently in Ref.~\cite{YU1}, albeit explicit results have been obtained only in the case of a clear circular aperture.

{\bfseries \em Acknowledgments.}---
We thank M.~Barbieri, P.~Boucher, D.~Braun, M.~G.~Genoni, M.~Guta, H.~Harmon, M.~Khalifa, P.~Knott, I.~Lesanovsky, P.~Liuzzo-Scorpo, C.~Lupo, R.~Nair, R.~Nichols, C.~Oh, M.~Pezzali, S.~Pirandola, S.~Ragy, R.~Su, N.~Treps, M.~Tsang, and J.-P.~Wolf for useful discussions. This work was supported by the European Research Council under the Starting Grant GQCOP (Grant No.~637352), the Royal Society under the International Exchanges Programme (Grant No.~IE150570), the EPSRC under a Manufacturing Fellowship (Grant No.~EP/M008983/1), the University of Nottingham under a Nottingham Research Fellowship and a FROG Scholarship, and the EPSRC DTG Centre in Complex Systems and Processes.
\end{acknowledgments}


\bibliographystyle{apsrevfixedwithtitles}
\bibliography{biblio_tqsrieaas}

\begin{thebibliography}{64}%
\makeatletter
\providecommand \@ifxundefined [1]{%
 \@ifx{#1\undefined}
}%
\providecommand \@ifnum [1]{%
 \ifnum #1\expandafter \@firstoftwo
 \else \expandafter \@secondoftwo
 \fi
}%
\providecommand \@ifx [1]{%
 \ifx #1\expandafter \@firstoftwo
 \else \expandafter \@secondoftwo
 \fi
}%
\providecommand \natexlab [1]{#1}%
\providecommand \enquote  [1]{``#1''}%
\providecommand \bibnamefont  [1]{#1}%
\providecommand \bibfnamefont [1]{#1}%
\providecommand \citenamefont [1]{#1}%
\providecommand \href@noop [0]{\@secondoftwo}%
\providecommand \href [0]{\begingroup \@sanitize@url \@href}%
\providecommand \@href[1]{\@@startlink{#1}\@@href}%
\providecommand \@@href[1]{\endgroup#1\@@endlink}%
\providecommand \@sanitize@url [0]{\catcode `\\12\catcode `\$12\catcode
  `\&12\catcode `\#12\catcode `\^12\catcode `\_12\catcode `\%12\relax}%
\providecommand \@@startlink[1]{}%
\providecommand \@@endlink[0]{}%
\providecommand \url  [0]{\begingroup\@sanitize@url \@url }%
\providecommand \@url [1]{\endgroup\@href {#1}{\urlprefix }}%
\providecommand \urlprefix  [0]{URL }%
\providecommand \Eprint [0]{\href }%
\providecommand \doibase [0]{http://dx.doi.org/}%
\providecommand \selectlanguage [0]{\@gobble}%
\providecommand \bibinfo  [0]{\@secondoftwo}%
\providecommand \bibfield  [0]{\@secondoftwo}%
\providecommand \translation [1]{[#1]}%
\providecommand \BibitemOpen [0]{}%
\providecommand \bibitemStop [0]{}%
\providecommand \bibitemNoStop [0]{.\EOS\space}%
\providecommand \EOS [0]{\spacefactor3000\relax}%
\providecommand \BibitemShut  [1]{\csname bibitem#1\endcsname}%
\let\auto@bib@innerbib\@empty
\bibitem [{\citenamefont {Rayleigh}(1879)}]{rayleigh1879xxxi}%
  \BibitemOpen
  \bibfield  {author} {\bibinfo {author} {\bibfnamefont {L.}~\bibnamefont
  {Rayleigh}},\ }\bibfield  {title} {\enquote {\bibinfo {title} {Xxxi.
  investigations in optics, with special reference to the spectroscope},}\
  }\href@noop {} {\bibfield  {journal} {\bibinfo  {journal} {The London,
  Edinburgh, and Dublin Philosophical Magazine and Journal of Science}\
  }\textbf {\bibinfo {volume} {8}},\ \bibinfo {pages} {261} (\bibinfo {year}
  {1879})}\BibitemShut {NoStop}%
\bibitem [{\citenamefont {Tsang}\ \emph {et~al.}(2016)\citenamefont {Tsang},
  \citenamefont {Nair},\ and\ \citenamefont {Lu}}]{tsang2016quantum}%
  \BibitemOpen
  \bibfield  {author} {\bibinfo {author} {\bibfnamefont {M.}~\bibnamefont
  {Tsang}}, \bibinfo {author} {\bibfnamefont {R.}~\bibnamefont {Nair}}, \ and\
  \bibinfo {author} {\bibfnamefont {X.-M.}\ \bibnamefont {Lu}},\ }\bibfield
  {title} {\enquote {\bibinfo {title} {Quantum theory of superresolution for
  two incoherent optical point sources},}\ }\href@noop {} {\bibfield  {journal}
  {\bibinfo  {journal} {Physical Review X}\ }\textbf {\bibinfo {volume} {6}},\
  \bibinfo {pages} {031033} (\bibinfo {year} {2016})}\BibitemShut {NoStop}%
\bibitem [{\citenamefont {M\"ockl}\ \emph {et~al.}()\citenamefont {M\"ockl},
  \citenamefont {Lamb},\ and\ \citenamefont {Bräuchle}}]{fluorescencenobel}%
  \BibitemOpen
  \bibfield  {author} {\bibinfo {author} {\bibfnamefont {L.}~\bibnamefont
  {M\"ockl}}, \bibinfo {author} {\bibfnamefont {D.~C.}\ \bibnamefont {Lamb}}, \
  and\ \bibinfo {author} {\bibfnamefont {C.}~\bibnamefont {Bräuchle}},\
  }\bibfield  {title} {\enquote {\bibinfo {title} {{Super-resolved Fluorescence
  Microscopy: Nobel Prize in Chemistry 2014 for Eric~Betzig, Stefan~Hell, and
  William~E.~Moerner}},}\ }\href {\doibase 10.1002/anie.201410265} {\bibfield
  {journal} {\bibinfo  {journal} {Angewandte Chemie International Edition}\
  }\textbf {\bibinfo {volume} {53}},\ \bibinfo {pages} {13972}}\BibitemShut
  {NoStop}%
\bibitem [{\citenamefont {Moerner}(2007)}]{moerner2007new}%
  \BibitemOpen
  \bibfield  {author} {\bibinfo {author} {\bibfnamefont {W.~E.}\ \bibnamefont
  {Moerner}},\ }\bibfield  {title} {\enquote {\bibinfo {title} {New directions
  in single-molecule imaging and analysis},}\ }\href {\doibase
  10.1073/pnas.0610081104} {\bibfield  {journal} {\bibinfo  {journal}
  {Proceedings of the National Academy of Sciences}\ }\textbf {\bibinfo
  {volume} {104}},\ \bibinfo {pages} {12596} (\bibinfo {year} {2007})},\
  \Eprint
  {http://arxiv.org/abs/http://www.pnas.org/content/104/31/12596.full.pdf}
  {http://www.pnas.org/content/104/31/12596.full.pdf} \BibitemShut {NoStop}%
\bibitem [{\citenamefont {Leach}\ and\ \citenamefont
  {Sherlock}(2014)}]{leach2014applications}%
  \BibitemOpen
  \bibfield  {author} {\bibinfo {author} {\bibfnamefont {R.}~\bibnamefont
  {Leach}}\ and\ \bibinfo {author} {\bibfnamefont {B.}~\bibnamefont
  {Sherlock}},\ }\bibfield  {title} {\enquote {\bibinfo {title} {Applications
  of super-resolution imaging in the field of surface topography
  measurement},}\ }\href {\doibase 10.1088/2051-672X/2/2/023001} {\bibfield
  {journal} {\bibinfo  {journal} {Surface Topography: Metrology and
  Properties}\ }\textbf {\bibinfo {volume} {2}},\ \bibinfo {pages} {023001}
  (\bibinfo {year} {2014})}\BibitemShut {NoStop}%
\bibitem [{\citenamefont {Genovese}(2016)}]{genovese2016real}%
  \BibitemOpen
  \bibfield  {author} {\bibinfo {author} {\bibfnamefont {M.}~\bibnamefont
  {Genovese}},\ }\bibfield  {title} {\enquote {\bibinfo {title} {Real
  applications of quantum imaging},}\ }\href {\doibase
  10.1088/2040-8978/18/7/073002} {\bibfield  {journal} {\bibinfo  {journal}
  {Journal of Optics}\ }\textbf {\bibinfo {volume} {18}},\ \bibinfo {pages}
  {073002} (\bibinfo {year} {2016})}\BibitemShut {NoStop}%
\bibitem [{\citenamefont {Nair}\ and\ \citenamefont
  {Tsang}(2016{\natexlab{a}})}]{nair2016interferometric}%
  \BibitemOpen
  \bibfield  {author} {\bibinfo {author} {\bibfnamefont {R.}~\bibnamefont
  {Nair}}\ and\ \bibinfo {author} {\bibfnamefont {M.}~\bibnamefont {Tsang}},\
  }\bibfield  {title} {\enquote {\bibinfo {title} {Interferometric
  superlocalization of two incoherent optical point sources},}\ }\href@noop {}
  {\bibfield  {journal} {\bibinfo  {journal} {Optics express}\ }\textbf
  {\bibinfo {volume} {24}},\ \bibinfo {pages} {3684} (\bibinfo {year}
  {2016}{\natexlab{a}})}\BibitemShut {NoStop}%
\bibitem [{\citenamefont {Nair}\ and\ \citenamefont
  {Tsang}(2016{\natexlab{b}})}]{nair2016far}%
  \BibitemOpen
  \bibfield  {author} {\bibinfo {author} {\bibfnamefont {R.}~\bibnamefont
  {Nair}}\ and\ \bibinfo {author} {\bibfnamefont {M.}~\bibnamefont {Tsang}},\
  }\bibfield  {title} {\enquote {\bibinfo {title} {Far-field superresolution of
  thermal electromagnetic sources at the quantum limit},}\ }\href {\doibase
  10.1103/PhysRevLett.117.190801} {\bibfield  {journal} {\bibinfo  {journal}
  {Physical Review Letters}\ }\textbf {\bibinfo {volume} {117}},\ \bibinfo
  {pages} {190801} (\bibinfo {year} {2016}{\natexlab{b}})}\BibitemShut
  {NoStop}%
\bibitem [{\citenamefont {Lupo}\ and\ \citenamefont
  {Pirandola}(2016)}]{lupo2016ultimate}%
  \BibitemOpen
  \bibfield  {author} {\bibinfo {author} {\bibfnamefont {C.}~\bibnamefont
  {Lupo}}\ and\ \bibinfo {author} {\bibfnamefont {S.}~\bibnamefont
  {Pirandola}},\ }\bibfield  {title} {\enquote {\bibinfo {title} {Ultimate
  precision bound of quantum and subwavelength imaging},}\ }\href {\doibase
  10.1103/PhysRevLett.117.190802} {\bibfield  {journal} {\bibinfo  {journal}
  {Physical Review Letters}\ }\textbf {\bibinfo {volume} {117}},\ \bibinfo
  {pages} {190802} (\bibinfo {year} {2016})}\BibitemShut {NoStop}%
\bibitem [{\citenamefont {Tsang}(2017)}]{tsang2017subdiffraction}%
  \BibitemOpen
  \bibfield  {author} {\bibinfo {author} {\bibfnamefont {M.}~\bibnamefont
  {Tsang}},\ }\bibfield  {title} {\enquote {\bibinfo {title} {Subdiffraction
  incoherent optical imaging via spatial-mode demultiplexing},}\ }\href@noop {}
  {\bibfield  {journal} {\bibinfo  {journal} {New Journal of Physics}\ }\textbf
  {\bibinfo {volume} {19}},\ \bibinfo {pages} {023054} (\bibinfo {year}
  {2017})}\BibitemShut {NoStop}%
\bibitem [{\citenamefont {Kerviche}\ \emph {et~al.}(2017)\citenamefont
  {Kerviche}, \citenamefont {Guha},\ and\ \citenamefont
  {Ashok}}]{kerviche2017fundamental}%
  \BibitemOpen
  \bibfield  {author} {\bibinfo {author} {\bibfnamefont {R.}~\bibnamefont
  {Kerviche}}, \bibinfo {author} {\bibfnamefont {S.}~\bibnamefont {Guha}}, \
  and\ \bibinfo {author} {\bibfnamefont {A.}~\bibnamefont {Ashok}},\ }\bibfield
   {title} {\enquote {\bibinfo {title} {Fundamental limit of resolving two
  point sources limited by an arbitrary point spread function},}\ }in\
  \href@noop {} {\emph {\bibinfo {booktitle} {Information Theory (ISIT), 2017
  IEEE International Symposium on}}}\ (\bibinfo {organization} {IEEE},\
  \bibinfo {year} {2017})\ pp.\ \bibinfo {pages} {441--445}\BibitemShut
  {NoStop}%
\bibitem [{\citenamefont {Ang}\ \emph {et~al.}(2017)\citenamefont {Ang},
  \citenamefont {Nair},\ and\ \citenamefont {Tsang}}]{ang2017quantum}%
  \BibitemOpen
  \bibfield  {author} {\bibinfo {author} {\bibfnamefont {S.~Z.}\ \bibnamefont
  {Ang}}, \bibinfo {author} {\bibfnamefont {R.}~\bibnamefont {Nair}}, \ and\
  \bibinfo {author} {\bibfnamefont {M.}~\bibnamefont {Tsang}},\ }\bibfield
  {title} {\enquote {\bibinfo {title} {Quantum limit for two-dimensional
  resolution of two incoherent optical point sources},}\ }\href {\doibase
  10.1103/PhysRevA.95.063847} {\bibfield  {journal} {\bibinfo  {journal}
  {Physical Review A}\ }\textbf {\bibinfo {volume} {95}},\ \bibinfo {pages}
  {063847} (\bibinfo {year} {2017})}\BibitemShut {NoStop}%
\bibitem [{\citenamefont {Chrostowski}\ \emph {et~al.}(2017)\citenamefont
  {Chrostowski}, \citenamefont {Demkowicz-Dobrza{\'n}ski}, \citenamefont
  {Jarzyna},\ and\ \citenamefont {Banaszek}}]{chrostowski2017super}%
  \BibitemOpen
  \bibfield  {author} {\bibinfo {author} {\bibfnamefont {A.}~\bibnamefont
  {Chrostowski}}, \bibinfo {author} {\bibfnamefont {R.}~\bibnamefont
  {Demkowicz-Dobrza{\'n}ski}}, \bibinfo {author} {\bibfnamefont
  {M.}~\bibnamefont {Jarzyna}}, \ and\ \bibinfo {author} {\bibfnamefont
  {K.}~\bibnamefont {Banaszek}},\ }\bibfield  {title} {\enquote {\bibinfo
  {title} {On super-resolution imaging as a multiparameter estimation
  problem},}\ }\href@noop {} {\bibfield  {journal} {\bibinfo  {journal}
  {International Journal of Quantum Information}\ ,\ \bibinfo {pages}
  {1740005}} (\bibinfo {year} {2017})}\BibitemShut {NoStop}%
\bibitem [{\citenamefont {{\v{R}}eha{\v{c}}ek}\ \emph
  {et~al.}(2017{\natexlab{a}})\citenamefont {{\v{R}}eha{\v{c}}ek},
  \citenamefont {Hradil}, \citenamefont {Stoklasa}, \citenamefont {Pa{\'u}r},
  \citenamefont {Grover}, \citenamefont {Krzic},\ and\ \citenamefont
  {S{\'a}nchez-Soto}}]{rehavcek2017multiparameter}%
  \BibitemOpen
  \bibfield  {author} {\bibinfo {author} {\bibfnamefont {J.}~\bibnamefont
  {{\v{R}}eha{\v{c}}ek}}, \bibinfo {author} {\bibfnamefont {Z.}~\bibnamefont
  {Hradil}}, \bibinfo {author} {\bibfnamefont {B.}~\bibnamefont {Stoklasa}},
  \bibinfo {author} {\bibfnamefont {M.}~\bibnamefont {Pa{\'u}r}}, \bibinfo
  {author} {\bibfnamefont {J.}~\bibnamefont {Grover}}, \bibinfo {author}
  {\bibfnamefont {A.}~\bibnamefont {Krzic}}, \ and\ \bibinfo {author}
  {\bibfnamefont {L.}~\bibnamefont {S{\'a}nchez-Soto}},\ }\bibfield  {title}
  {\enquote {\bibinfo {title} {Multiparameter quantum metrology of incoherent
  point sources: Towards realistic superresolution},}\ }\href@noop {}
  {\bibfield  {journal} {\bibinfo  {journal} {Physical Review A}\ }\textbf
  {\bibinfo {volume} {96}},\ \bibinfo {pages} {062107} (\bibinfo {year}
  {2017}{\natexlab{a}})}\BibitemShut {NoStop}%
\bibitem [{\citenamefont {{\v{R}}eha{\v{c}}ek}\ \emph
  {et~al.}(2017{\natexlab{b}})\citenamefont {{\v{R}}eha{\v{c}}ek},
  \citenamefont {Pa{\'u}r}, \citenamefont {Stoklasa}, \citenamefont {Hradil},\
  and\ \citenamefont {S{\'a}nchez-Soto}}]{rehacek2017optimal}%
  \BibitemOpen
  \bibfield  {author} {\bibinfo {author} {\bibfnamefont {J.}~\bibnamefont
  {{\v{R}}eha{\v{c}}ek}}, \bibinfo {author} {\bibfnamefont {M.}~\bibnamefont
  {Pa{\'u}r}}, \bibinfo {author} {\bibfnamefont {B.}~\bibnamefont {Stoklasa}},
  \bibinfo {author} {\bibfnamefont {Z.}~\bibnamefont {Hradil}}, \ and\ \bibinfo
  {author} {\bibfnamefont {L.}~\bibnamefont {S{\'a}nchez-Soto}},\ }\bibfield
  {title} {\enquote {\bibinfo {title} {Optimal measurements for resolution
  beyond the rayleigh limit},}\ }\href@noop {} {\bibfield  {journal} {\bibinfo
  {journal} {Optics Letters}\ }\textbf {\bibinfo {volume} {42}},\ \bibinfo
  {pages} {231} (\bibinfo {year} {2017}{\natexlab{b}})}\BibitemShut {NoStop}%
\bibitem [{\citenamefont {{\v{R}}eh\'a{\v{c}}ek}\ \emph
  {et~al.}(2018)\citenamefont {{\v{R}}eh\'a{\v{c}}ek}, \citenamefont {Hradil},
  \citenamefont {Koutn\'y}, \citenamefont {Grover}, \citenamefont {Krzic},\
  and\ \citenamefont {S\'anchez-Soto}}]{rehacek2017optimal2}%
  \BibitemOpen
  \bibfield  {author} {\bibinfo {author} {\bibfnamefont {J.}~\bibnamefont
  {{\v{R}}eh\'a{\v{c}}ek}}, \bibinfo {author} {\bibfnamefont {Z.}~\bibnamefont
  {Hradil}}, \bibinfo {author} {\bibfnamefont {D.}~\bibnamefont {Koutn\'y}},
  \bibinfo {author} {\bibfnamefont {J.}~\bibnamefont {Grover}}, \bibinfo
  {author} {\bibfnamefont {A.}~\bibnamefont {Krzic}}, \ and\ \bibinfo {author}
  {\bibfnamefont {L.~L.}\ \bibnamefont {S\'anchez-Soto}},\ }\bibfield  {title}
  {\enquote {\bibinfo {title} {Optimal measurements for quantum spatial
  superresolution},}\ }\href@noop {} {\bibfield  {journal} {\bibinfo  {journal}
  {Physical Review A}\ }\textbf {\bibinfo {volume} {98}},\ \bibinfo {pages}
  {012103} (\bibinfo {year} {2018})}\BibitemShut {NoStop}%
\bibitem [{\citenamefont {Zhou}\ and\ \citenamefont
  {Jiang}(2019)}]{zhou2018modern}%
  \BibitemOpen
  \bibfield  {author} {\bibinfo {author} {\bibfnamefont {S.}~\bibnamefont
  {Zhou}}\ and\ \bibinfo {author} {\bibfnamefont {L.}~\bibnamefont {Jiang}},\
  }\bibfield  {title} {\enquote {\bibinfo {title} {Modern description of
  rayleigh's criterion},}\ }\href@noop {} {\bibfield  {journal} {\bibinfo
  {journal} {Physical Review A}\ }\textbf {\bibinfo {volume} {99}},\ \bibinfo
  {pages} {013808} (\bibinfo {year} {2019})}\BibitemShut {NoStop}%
\bibitem [{\citenamefont {Tsang}(2019)}]{tsang2019quantum}%
  \BibitemOpen
  \bibfield  {author} {\bibinfo {author} {\bibfnamefont {M.}~\bibnamefont
  {Tsang}},\ }\bibfield  {title} {\enquote {\bibinfo {title} {Quantum limit to
  subdiffraction incoherent optical imaging},}\ }\href@noop {} {\bibfield
  {journal} {\bibinfo  {journal} {Physical Review A}\ }\textbf {\bibinfo
  {volume} {99}},\ \bibinfo {pages} {012305} (\bibinfo {year}
  {2019})}\BibitemShut {NoStop}%
\bibitem [{\citenamefont {Helstrom}(1976)}]{helstrom1976quantum}%
  \BibitemOpen
  \bibfield  {author} {\bibinfo {author} {\bibfnamefont {C.~W.}\ \bibnamefont
  {Helstrom}},\ }\href@noop {} {\emph {\bibinfo {title} {{Quantum Detection and
  Estimation Theory}}}}\ (\bibinfo  {publisher} {Academic Press},\ \bibinfo
  {address} {New York},\ \bibinfo {year} {1976})\BibitemShut {NoStop}%
\bibitem [{\citenamefont {Braunstein}\ and\ \citenamefont
  {Caves}(1994)}]{braunstein1994statistical}%
  \BibitemOpen
  \bibfield  {author} {\bibinfo {author} {\bibfnamefont {S.~L.}\ \bibnamefont
  {Braunstein}}\ and\ \bibinfo {author} {\bibfnamefont {C.~M.}\ \bibnamefont
  {Caves}},\ }\bibfield  {title} {\enquote {\bibinfo {title} {Statistical
  distance and the geometry of quantum states},}\ }\href {\doibase
  10.1103/PhysRevLett.72.3439} {\bibfield  {journal} {\bibinfo  {journal}
  {Physical Review Letters}\ }\textbf {\bibinfo {volume} {72}},\ \bibinfo
  {pages} {3439} (\bibinfo {year} {1994})}\BibitemShut {NoStop}%
\bibitem [{\citenamefont {Paris}(2009)}]{paris2009quantum}%
  \BibitemOpen
  \bibfield  {author} {\bibinfo {author} {\bibfnamefont {M.~G.~A.}\
  \bibnamefont {Paris}},\ }\bibfield  {title} {\enquote {\bibinfo {title}
  {Quantum estimation for quantum technology},}\ }\href {\doibase
  10.1142/S0219749909004839} {\bibfield  {journal} {\bibinfo  {journal}
  {International Journal of Quantum Information}\ }\textbf {\bibinfo {volume}
  {07}},\ \bibinfo {pages} {125} (\bibinfo {year} {2009})}\BibitemShut
  {NoStop}%
\bibitem [{\citenamefont {Giovannetti}\ \emph {et~al.}(2011)\citenamefont
  {Giovannetti}, \citenamefont {Lloyd},\ and\ \citenamefont
  {Maccone}}]{giovannetti2011advances}%
  \BibitemOpen
  \bibfield  {author} {\bibinfo {author} {\bibfnamefont {V.}~\bibnamefont
  {Giovannetti}}, \bibinfo {author} {\bibfnamefont {S.}~\bibnamefont {Lloyd}},
  \ and\ \bibinfo {author} {\bibfnamefont {L.}~\bibnamefont {Maccone}},\
  }\bibfield  {title} {\enquote {\bibinfo {title} {Advances in quantum
  metrology},}\ }\href@noop {} {\bibfield  {journal} {\bibinfo  {journal}
  {Nature Photonics}\ }\textbf {\bibinfo {volume} {5}},\ \bibinfo {pages} {222}
  (\bibinfo {year} {2011})}\BibitemShut {NoStop}%
\bibitem [{\citenamefont {Yang}\ \emph {et~al.}(2016)\citenamefont {Yang},
  \citenamefont {Tashchilina}, \citenamefont {Moiseev}, \citenamefont {Simon},\
  and\ \citenamefont {Lvovsky}}]{yang2016far}%
  \BibitemOpen
  \bibfield  {author} {\bibinfo {author} {\bibfnamefont {F.}~\bibnamefont
  {Yang}}, \bibinfo {author} {\bibfnamefont {A.}~\bibnamefont {Tashchilina}},
  \bibinfo {author} {\bibfnamefont {E.~S.}\ \bibnamefont {Moiseev}}, \bibinfo
  {author} {\bibfnamefont {C.}~\bibnamefont {Simon}}, \ and\ \bibinfo {author}
  {\bibfnamefont {A.~I.}\ \bibnamefont {Lvovsky}},\ }\bibfield  {title}
  {\enquote {\bibinfo {title} {Far-field linear optical superresolution via
  heterodyne detection in a higher-order local oscillator mode},}\ }\href
  {\doibase 10.1364/OPTICA.3.001148} {\bibfield  {journal} {\bibinfo  {journal}
  {Optica}\ }\textbf {\bibinfo {volume} {3}},\ \bibinfo {pages} {1148}
  (\bibinfo {year} {2016})}\BibitemShut {NoStop}%
\bibitem [{\citenamefont {Pa\'{u}r}\ \emph {et~al.}(2016)\citenamefont
  {Pa\'{u}r}, \citenamefont {Stoklasa}, \citenamefont {Hradil}, \citenamefont
  {S\'{a}nchez-Soto},\ and\ \citenamefont {Rehacek}}]{paur2016achieving}%
  \BibitemOpen
  \bibfield  {author} {\bibinfo {author} {\bibfnamefont {M.}~\bibnamefont
  {Pa\'{u}r}}, \bibinfo {author} {\bibfnamefont {B.}~\bibnamefont {Stoklasa}},
  \bibinfo {author} {\bibfnamefont {Z.}~\bibnamefont {Hradil}}, \bibinfo
  {author} {\bibfnamefont {L.~L.}\ \bibnamefont {S\'{a}nchez-Soto}}, \ and\
  \bibinfo {author} {\bibfnamefont {J.}~\bibnamefont {Rehacek}},\ }\bibfield
  {title} {\enquote {\bibinfo {title} {Achieving the ultimate optical
  resolution},}\ }\href {\doibase 10.1364/OPTICA.3.001144} {\bibfield
  {journal} {\bibinfo  {journal} {Optica}\ }\textbf {\bibinfo {volume} {3}},\
  \bibinfo {pages} {1144} (\bibinfo {year} {2016})}\BibitemShut {NoStop}%
\bibitem [{\citenamefont {Tang}\ \emph {et~al.}(2016)\citenamefont {Tang},
  \citenamefont {Durak},\ and\ \citenamefont {Ling}}]{tang2016fault}%
  \BibitemOpen
  \bibfield  {author} {\bibinfo {author} {\bibfnamefont {Z.~S.}\ \bibnamefont
  {Tang}}, \bibinfo {author} {\bibfnamefont {K.}~\bibnamefont {Durak}}, \ and\
  \bibinfo {author} {\bibfnamefont {A.}~\bibnamefont {Ling}},\ }\bibfield
  {title} {\enquote {\bibinfo {title} {Fault-tolerant and finite-error
  localization for point emitters within the diffraction limit},}\ }in\ \href
  {\doibase 10.1364/FIO.2016.FTu2F.6} {\emph {\bibinfo {booktitle} {Frontiers
  in Optics 2016}}}\ (\bibinfo  {publisher} {Optical Society of America},\
  \bibinfo {year} {2016})\ p.\ \bibinfo {pages} {FTu2F.6}\BibitemShut {NoStop}%
\bibitem [{\citenamefont {Tham}\ \emph {et~al.}(2017)\citenamefont {Tham},
  \citenamefont {Ferretti},\ and\ \citenamefont {Steinberg}}]{tham2017beating}%
  \BibitemOpen
  \bibfield  {author} {\bibinfo {author} {\bibfnamefont {W.-K.}\ \bibnamefont
  {Tham}}, \bibinfo {author} {\bibfnamefont {H.}~\bibnamefont {Ferretti}}, \
  and\ \bibinfo {author} {\bibfnamefont {A.~M.}\ \bibnamefont {Steinberg}},\
  }\bibfield  {title} {\enquote {\bibinfo {title} {Beating rayleigh's curse by
  imaging using phase information},}\ }\href {\doibase
  10.1103/PhysRevLett.118.070801} {\bibfield  {journal} {\bibinfo  {journal}
  {Physical Review Letters}\ }\textbf {\bibinfo {volume} {118}},\ \bibinfo
  {pages} {070801} (\bibinfo {year} {2017})}\BibitemShut {NoStop}%
\bibitem [{\citenamefont {Leach}\ \emph {et~al.}(2015)\citenamefont {Leach},
  \citenamefont {Evans}, \citenamefont {He}, \citenamefont {Davies},
  \citenamefont {Duparré}, \citenamefont {Henning}, \citenamefont {Jones},\
  and\ \citenamefont {O'Connor}}]{leach2015open}%
  \BibitemOpen
  \bibfield  {author} {\bibinfo {author} {\bibfnamefont {R.}~\bibnamefont
  {Leach}}, \bibinfo {author} {\bibfnamefont {C.}~\bibnamefont {Evans}},
  \bibinfo {author} {\bibfnamefont {L.}~\bibnamefont {He}}, \bibinfo {author}
  {\bibfnamefont {A.}~\bibnamefont {Davies}}, \bibinfo {author} {\bibfnamefont
  {A.}~\bibnamefont {Duparré}}, \bibinfo {author} {\bibfnamefont
  {A.}~\bibnamefont {Henning}}, \bibinfo {author} {\bibfnamefont {C.~W.}\
  \bibnamefont {Jones}}, \ and\ \bibinfo {author} {\bibfnamefont
  {D.}~\bibnamefont {O'Connor}},\ }\bibfield  {title} {\enquote {\bibinfo
  {title} {Open questions in surface topography measurement: a roadmap},}\
  }\href {\doibase 10.1088/2051-672X/3/1/013001} {\bibfield  {journal}
  {\bibinfo  {journal} {Surface Topography: Metrology and Properties}\ }\textbf
  {\bibinfo {volume} {3}},\ \bibinfo {pages} {013001} (\bibinfo {year}
  {2015})}\BibitemShut {NoStop}%
\bibitem [{\citenamefont {De~Lega}\ and\ \citenamefont
  {de~Groot}(2012)}]{de2012lateral}%
  \BibitemOpen
  \bibfield  {author} {\bibinfo {author} {\bibfnamefont {X.~C.}\ \bibnamefont
  {De~Lega}}\ and\ \bibinfo {author} {\bibfnamefont {P.}~\bibnamefont
  {de~Groot}},\ }\bibfield  {title} {\enquote {\bibinfo {title} {Lateral
  resolution and instrument transfer function as criteria for selecting surface
  metrology instruments},}\ }in\ \href@noop {} {\emph {\bibinfo {booktitle}
  {Optical Fabrication and Testing}}}\ (\bibinfo {organization} {Optical
  Society of America},\ \bibinfo {year} {2012})\ pp.\ \bibinfo {pages}
  {OTu1D--4}\BibitemShut {NoStop}%
\bibitem [{\citenamefont {Matsumoto}(2002)}]{matsumoto2002new}%
  \BibitemOpen
  \bibfield  {author} {\bibinfo {author} {\bibfnamefont {K.}~\bibnamefont
  {Matsumoto}},\ }\bibfield  {title} {\enquote {\bibinfo {title} {A new
  approach to the cram{\'e}r-rao-type bound of the pure-state model},}\
  }\href@noop {} {\bibfield  {journal} {\bibinfo  {journal} {Journal of Physics
  A: Mathematical and General}\ }\textbf {\bibinfo {volume} {35}},\ \bibinfo
  {pages} {3111} (\bibinfo {year} {2002})}\BibitemShut {NoStop}%
\bibitem [{\citenamefont {T{\'o}th}\ and\ \citenamefont
  {Apellaniz}(2014)}]{toth2014quantum}%
  \BibitemOpen
  \bibfield  {author} {\bibinfo {author} {\bibfnamefont {G.}~\bibnamefont
  {T{\'o}th}}\ and\ \bibinfo {author} {\bibfnamefont {I.}~\bibnamefont
  {Apellaniz}},\ }\bibfield  {title} {\enquote {\bibinfo {title} {Quantum
  metrology from a quantum information science perspective},}\ }\href@noop {}
  {\bibfield  {journal} {\bibinfo  {journal} {Journal of Physics A:
  Mathematical and Theoretical}\ }\textbf {\bibinfo {volume} {47}},\ \bibinfo
  {pages} {424006} (\bibinfo {year} {2014})}\BibitemShut {NoStop}%
\bibitem [{\citenamefont {Szczykulska}\ \emph {et~al.}(2016)\citenamefont
  {Szczykulska}, \citenamefont {Baumgratz},\ and\ \citenamefont
  {Datta}}]{szczykulska2016multi}%
  \BibitemOpen
  \bibfield  {author} {\bibinfo {author} {\bibfnamefont {M.}~\bibnamefont
  {Szczykulska}}, \bibinfo {author} {\bibfnamefont {T.}~\bibnamefont
  {Baumgratz}}, \ and\ \bibinfo {author} {\bibfnamefont {A.}~\bibnamefont
  {Datta}},\ }\bibfield  {title} {\enquote {\bibinfo {title} {Multi-parameter
  quantum metrology},}\ }\href@noop {} {\bibfield  {journal} {\bibinfo
  {journal} {Advances in Physics: X}\ }\textbf {\bibinfo {volume} {1}},\
  \bibinfo {pages} {621} (\bibinfo {year} {2016})}\BibitemShut {NoStop}%
\bibitem [{\citenamefont {Ragy}\ \emph {et~al.}(2016)\citenamefont {Ragy},
  \citenamefont {Jarzyna},\ and\ \citenamefont
  {Demkowicz-Dobrza{\'n}ski}}]{ragy2016compatibility}%
  \BibitemOpen
  \bibfield  {author} {\bibinfo {author} {\bibfnamefont {S.}~\bibnamefont
  {Ragy}}, \bibinfo {author} {\bibfnamefont {M.}~\bibnamefont {Jarzyna}}, \
  and\ \bibinfo {author} {\bibfnamefont {R.}~\bibnamefont
  {Demkowicz-Dobrza{\'n}ski}},\ }\bibfield  {title} {\enquote {\bibinfo {title}
  {Compatibility in multiparameter quantum metrology},}\ }\href@noop {}
  {\bibfield  {journal} {\bibinfo  {journal} {Physical Review A}\ }\textbf
  {\bibinfo {volume} {94}},\ \bibinfo {pages} {052108} (\bibinfo {year}
  {2016})}\BibitemShut {NoStop}%
\bibitem [{\citenamefont {Braun}\ \emph {et~al.}(2018)\citenamefont {Braun},
  \citenamefont {Adesso}, \citenamefont {Benatti}, \citenamefont {Floreanini},
  \citenamefont {Marzolino}, \citenamefont {Mitchell},\ and\ \citenamefont
  {Pirandola}}]{braun2017quantum}%
  \BibitemOpen
  \bibfield  {author} {\bibinfo {author} {\bibfnamefont {D.}~\bibnamefont
  {Braun}}, \bibinfo {author} {\bibfnamefont {G.}~\bibnamefont {Adesso}},
  \bibinfo {author} {\bibfnamefont {F.}~\bibnamefont {Benatti}}, \bibinfo
  {author} {\bibfnamefont {R.}~\bibnamefont {Floreanini}}, \bibinfo {author}
  {\bibfnamefont {U.}~\bibnamefont {Marzolino}}, \bibinfo {author}
  {\bibfnamefont {M.~W.}\ \bibnamefont {Mitchell}}, \ and\ \bibinfo {author}
  {\bibfnamefont {S.}~\bibnamefont {Pirandola}},\ }\bibfield  {title} {\enquote
  {\bibinfo {title} {Quantum-enhanced measurements without entanglement},}\
  }\href@noop {} {\bibfield  {journal} {\bibinfo  {journal} {Reviews of Modern
  Physics}\ }\textbf {\bibinfo {volume} {90}},\ \bibinfo {pages} {035006}
  (\bibinfo {year} {2018})}\BibitemShut {NoStop}%
\bibitem [{\citenamefont {Pezz\`e}\ \emph {et~al.}(2018)\citenamefont
  {Pezz\`e}, \citenamefont {Smerzi}, \citenamefont {Oberthaler}, \citenamefont
  {Schmied},\ and\ \citenamefont {Treutlein}}]{pezze2016quantum}%
  \BibitemOpen
  \bibfield  {author} {\bibinfo {author} {\bibfnamefont {L.}~\bibnamefont
  {Pezz\`e}}, \bibinfo {author} {\bibfnamefont {A.}~\bibnamefont {Smerzi}},
  \bibinfo {author} {\bibfnamefont {M.~K.}\ \bibnamefont {Oberthaler}},
  \bibinfo {author} {\bibfnamefont {R.}~\bibnamefont {Schmied}}, \ and\
  \bibinfo {author} {\bibfnamefont {P.}~\bibnamefont {Treutlein}},\ }\bibfield
  {title} {\enquote {\bibinfo {title} {Quantum metrology with nonclassical
  states of atomic ensembles},}\ }\href@noop {} {\bibfield  {journal} {\bibinfo
   {journal} {Reviews of Modern Physics}\ }\textbf {\bibinfo {volume} {90}},\
  \bibinfo {pages} {035005} (\bibinfo {year} {2018})}\BibitemShut {NoStop}%
\bibitem [{\citenamefont {Monras}\ and\ \citenamefont
  {Illuminati}(2011)}]{monras2011measurement}%
  \BibitemOpen
  \bibfield  {author} {\bibinfo {author} {\bibfnamefont {A.}~\bibnamefont
  {Monras}}\ and\ \bibinfo {author} {\bibfnamefont {F.}~\bibnamefont
  {Illuminati}},\ }\bibfield  {title} {\enquote {\bibinfo {title} {Measurement
  of damping and temperature: Precision bounds in gaussian dissipative
  channels},}\ }\href@noop {} {\bibfield  {journal} {\bibinfo  {journal}
  {Physical Review A}\ }\textbf {\bibinfo {volume} {83}},\ \bibinfo {pages}
  {012315} (\bibinfo {year} {2011})}\BibitemShut {NoStop}%
\bibitem [{\citenamefont {Genoni}\ \emph {et~al.}(2013)\citenamefont {Genoni},
  \citenamefont {Paris}, \citenamefont {Adesso}, \citenamefont {Nha},
  \citenamefont {Knight},\ and\ \citenamefont {Kim}}]{genoni2013optimal}%
  \BibitemOpen
  \bibfield  {author} {\bibinfo {author} {\bibfnamefont {M.}~\bibnamefont
  {Genoni}}, \bibinfo {author} {\bibfnamefont {M.}~\bibnamefont {Paris}},
  \bibinfo {author} {\bibfnamefont {G.}~\bibnamefont {Adesso}}, \bibinfo
  {author} {\bibfnamefont {H.}~\bibnamefont {Nha}}, \bibinfo {author}
  {\bibfnamefont {P.}~\bibnamefont {Knight}}, \ and\ \bibinfo {author}
  {\bibfnamefont {M.}~\bibnamefont {Kim}},\ }\bibfield  {title} {\enquote
  {\bibinfo {title} {Optimal estimation of joint parameters in phase space},}\
  }\href@noop {} {\bibfield  {journal} {\bibinfo  {journal} {Physical Review
  A}\ }\textbf {\bibinfo {volume} {87}},\ \bibinfo {pages} {012107} (\bibinfo
  {year} {2013})}\BibitemShut {NoStop}%
\bibitem [{\citenamefont {Steinlechner}\ \emph {et~al.}(2013)\citenamefont
  {Steinlechner}, \citenamefont {Bauchrowitz}, \citenamefont {Meinders},
  \citenamefont {M{\"u}ller-Ebhardt}, \citenamefont {Danzmann},\ and\
  \citenamefont {Schnabel}}]{steinlechner2013quantum}%
  \BibitemOpen
  \bibfield  {author} {\bibinfo {author} {\bibfnamefont {S.}~\bibnamefont
  {Steinlechner}}, \bibinfo {author} {\bibfnamefont {J.}~\bibnamefont
  {Bauchrowitz}}, \bibinfo {author} {\bibfnamefont {M.}~\bibnamefont
  {Meinders}}, \bibinfo {author} {\bibfnamefont {H.}~\bibnamefont
  {M{\"u}ller-Ebhardt}}, \bibinfo {author} {\bibfnamefont {K.}~\bibnamefont
  {Danzmann}}, \ and\ \bibinfo {author} {\bibfnamefont {R.}~\bibnamefont
  {Schnabel}},\ }\bibfield  {title} {\enquote {\bibinfo {title} {Quantum-dense
  metrology},}\ }\href@noop {} {\bibfield  {journal} {\bibinfo  {journal}
  {Nature Photonics}\ }\textbf {\bibinfo {volume} {7}},\ \bibinfo {pages} {626}
  (\bibinfo {year} {2013})}\BibitemShut {NoStop}%
\bibitem [{\citenamefont {Pinel}\ \emph {et~al.}(2013)\citenamefont {Pinel},
  \citenamefont {Jian}, \citenamefont {Treps}, \citenamefont {Fabre},\ and\
  \citenamefont {Braun}}]{pinel2013quantum}%
  \BibitemOpen
  \bibfield  {author} {\bibinfo {author} {\bibfnamefont {O.}~\bibnamefont
  {Pinel}}, \bibinfo {author} {\bibfnamefont {P.}~\bibnamefont {Jian}},
  \bibinfo {author} {\bibfnamefont {N.}~\bibnamefont {Treps}}, \bibinfo
  {author} {\bibfnamefont {C.}~\bibnamefont {Fabre}}, \ and\ \bibinfo {author}
  {\bibfnamefont {D.}~\bibnamefont {Braun}},\ }\bibfield  {title} {\enquote
  {\bibinfo {title} {Quantum parameter estimation using general single-mode
  gaussian states},}\ }\href@noop {} {\bibfield  {journal} {\bibinfo  {journal}
  {Physical Review A}\ }\textbf {\bibinfo {volume} {88}},\ \bibinfo {pages}
  {040102} (\bibinfo {year} {2013})}\BibitemShut {NoStop}%
\bibitem [{\citenamefont {Humphreys}\ \emph {et~al.}(2013)\citenamefont
  {Humphreys}, \citenamefont {Barbieri}, \citenamefont {Datta},\ and\
  \citenamefont {Walmsley}}]{humphreys2013quantum}%
  \BibitemOpen
  \bibfield  {author} {\bibinfo {author} {\bibfnamefont {P.~C.}\ \bibnamefont
  {Humphreys}}, \bibinfo {author} {\bibfnamefont {M.}~\bibnamefont {Barbieri}},
  \bibinfo {author} {\bibfnamefont {A.}~\bibnamefont {Datta}}, \ and\ \bibinfo
  {author} {\bibfnamefont {I.~A.}\ \bibnamefont {Walmsley}},\ }\bibfield
  {title} {\enquote {\bibinfo {title} {Quantum enhanced multiple phase
  estimation},}\ }\href@noop {} {\bibfield  {journal} {\bibinfo  {journal}
  {Physical Review Letters}\ }\textbf {\bibinfo {volume} {111}},\ \bibinfo
  {pages} {070403} (\bibinfo {year} {2013})}\BibitemShut {NoStop}%
\bibitem [{\citenamefont {Crowley}\ \emph {et~al.}(2014)\citenamefont
  {Crowley}, \citenamefont {Datta}, \citenamefont {Barbieri},\ and\
  \citenamefont {Walmsley}}]{crowley2014tradeoff}%
  \BibitemOpen
  \bibfield  {author} {\bibinfo {author} {\bibfnamefont {P.~J.}\ \bibnamefont
  {Crowley}}, \bibinfo {author} {\bibfnamefont {A.}~\bibnamefont {Datta}},
  \bibinfo {author} {\bibfnamefont {M.}~\bibnamefont {Barbieri}}, \ and\
  \bibinfo {author} {\bibfnamefont {I.~A.}\ \bibnamefont {Walmsley}},\
  }\bibfield  {title} {\enquote {\bibinfo {title} {Tradeoff in simultaneous
  quantum-limited phase and loss estimation in interferometry},}\ }\href@noop
  {} {\bibfield  {journal} {\bibinfo  {journal} {Physical Review A}\ }\textbf
  {\bibinfo {volume} {89}},\ \bibinfo {pages} {023845} (\bibinfo {year}
  {2014})}\BibitemShut {NoStop}%
\bibitem [{\citenamefont {Vidrighin}\ \emph {et~al.}(2014)\citenamefont
  {Vidrighin}, \citenamefont {Donati}, \citenamefont {Genoni}, \citenamefont
  {Jin}, \citenamefont {Kolthammer}, \citenamefont {Kim}, \citenamefont
  {Datta}, \citenamefont {Barbieri},\ and\ \citenamefont
  {Walmsley}}]{vidrighin2014joint}%
  \BibitemOpen
  \bibfield  {author} {\bibinfo {author} {\bibfnamefont {M.~D.}\ \bibnamefont
  {Vidrighin}}, \bibinfo {author} {\bibfnamefont {G.}~\bibnamefont {Donati}},
  \bibinfo {author} {\bibfnamefont {M.~G.}\ \bibnamefont {Genoni}}, \bibinfo
  {author} {\bibfnamefont {X.-M.}\ \bibnamefont {Jin}}, \bibinfo {author}
  {\bibfnamefont {W.~S.}\ \bibnamefont {Kolthammer}}, \bibinfo {author}
  {\bibfnamefont {M.}~\bibnamefont {Kim}}, \bibinfo {author} {\bibfnamefont
  {A.}~\bibnamefont {Datta}}, \bibinfo {author} {\bibfnamefont
  {M.}~\bibnamefont {Barbieri}}, \ and\ \bibinfo {author} {\bibfnamefont
  {I.~A.}\ \bibnamefont {Walmsley}},\ }\bibfield  {title} {\enquote {\bibinfo
  {title} {Joint estimation of phase and phase diffusion for quantum
  metrology},}\ }\href@noop {} {\bibfield  {journal} {\bibinfo  {journal}
  {Nature Communications}\ }\textbf {\bibinfo {volume} {5}},\ \bibinfo {pages}
  {3532} (\bibinfo {year} {2014})}\BibitemShut {NoStop}%
\bibitem [{\citenamefont {Banchi}\ \emph {et~al.}(2015)\citenamefont {Banchi},
  \citenamefont {Braunstein},\ and\ \citenamefont
  {Pirandola}}]{banchi2015quantum}%
  \BibitemOpen
  \bibfield  {author} {\bibinfo {author} {\bibfnamefont {L.}~\bibnamefont
  {Banchi}}, \bibinfo {author} {\bibfnamefont {S.~L.}\ \bibnamefont
  {Braunstein}}, \ and\ \bibinfo {author} {\bibfnamefont {S.}~\bibnamefont
  {Pirandola}},\ }\bibfield  {title} {\enquote {\bibinfo {title} {Quantum
  fidelity for arbitrary gaussian states},}\ }\href@noop {} {\bibfield
  {journal} {\bibinfo  {journal} {Physical Review Letters}\ }\textbf {\bibinfo
  {volume} {115}},\ \bibinfo {pages} {260501} (\bibinfo {year}
  {2015})}\BibitemShut {NoStop}%
\bibitem [{\citenamefont {Baumgratz}\ and\ \citenamefont
  {Datta}(2016)}]{baumgratz2016quantum}%
  \BibitemOpen
  \bibfield  {author} {\bibinfo {author} {\bibfnamefont {T.}~\bibnamefont
  {Baumgratz}}\ and\ \bibinfo {author} {\bibfnamefont {A.}~\bibnamefont
  {Datta}},\ }\bibfield  {title} {\enquote {\bibinfo {title} {Quantum enhanced
  estimation of a multidimensional field},}\ }\href@noop {} {\bibfield
  {journal} {\bibinfo  {journal} {Physical Review Letters}\ }\textbf {\bibinfo
  {volume} {116}},\ \bibinfo {pages} {030801} (\bibinfo {year}
  {2016})}\BibitemShut {NoStop}%
\bibitem [{\citenamefont {Altorio}\ \emph {et~al.}(2016)\citenamefont
  {Altorio}, \citenamefont {Genoni}, \citenamefont {Somma},\ and\ \citenamefont
  {Barbieri}}]{altorio2016metrology}%
  \BibitemOpen
  \bibfield  {author} {\bibinfo {author} {\bibfnamefont {M.}~\bibnamefont
  {Altorio}}, \bibinfo {author} {\bibfnamefont {M.~G.}\ \bibnamefont {Genoni}},
  \bibinfo {author} {\bibfnamefont {F.}~\bibnamefont {Somma}}, \ and\ \bibinfo
  {author} {\bibfnamefont {M.}~\bibnamefont {Barbieri}},\ }\bibfield  {title}
  {\enquote {\bibinfo {title} {Metrology with unknown detectors},}\ }\href@noop
  {} {\bibfield  {journal} {\bibinfo  {journal} {Physical Review Letters}\
  }\textbf {\bibinfo {volume} {116}},\ \bibinfo {pages} {100802} (\bibinfo
  {year} {2016})}\BibitemShut {NoStop}%
\bibitem [{\citenamefont {Pirandola}\ and\ \citenamefont
  {Lupo}(2017)}]{pirandola2017ultimate}%
  \BibitemOpen
  \bibfield  {author} {\bibinfo {author} {\bibfnamefont {S.}~\bibnamefont
  {Pirandola}}\ and\ \bibinfo {author} {\bibfnamefont {C.}~\bibnamefont
  {Lupo}},\ }\bibfield  {title} {\enquote {\bibinfo {title} {Ultimate precision
  of adaptive noise estimation},}\ }\href@noop {} {\bibfield  {journal}
  {\bibinfo  {journal} {Physical review letters}\ }\textbf {\bibinfo {volume}
  {118}},\ \bibinfo {pages} {100502} (\bibinfo {year} {2017})}\BibitemShut
  {NoStop}%
\bibitem [{\citenamefont {Pezz{\`e}}\ \emph {et~al.}(2017)\citenamefont
  {Pezz{\`e}}, \citenamefont {Ciampini}, \citenamefont {Spagnolo},
  \citenamefont {Humphreys}, \citenamefont {Datta}, \citenamefont {Walmsley},
  \citenamefont {Barbieri}, \citenamefont {Sciarrino},\ and\ \citenamefont
  {Smerzi}}]{pezze2017optimal}%
  \BibitemOpen
  \bibfield  {author} {\bibinfo {author} {\bibfnamefont {L.}~\bibnamefont
  {Pezz{\`e}}}, \bibinfo {author} {\bibfnamefont {M.~A.}\ \bibnamefont
  {Ciampini}}, \bibinfo {author} {\bibfnamefont {N.}~\bibnamefont {Spagnolo}},
  \bibinfo {author} {\bibfnamefont {P.~C.}\ \bibnamefont {Humphreys}}, \bibinfo
  {author} {\bibfnamefont {A.}~\bibnamefont {Datta}}, \bibinfo {author}
  {\bibfnamefont {I.~A.}\ \bibnamefont {Walmsley}}, \bibinfo {author}
  {\bibfnamefont {M.}~\bibnamefont {Barbieri}}, \bibinfo {author}
  {\bibfnamefont {F.}~\bibnamefont {Sciarrino}}, \ and\ \bibinfo {author}
  {\bibfnamefont {A.}~\bibnamefont {Smerzi}},\ }\bibfield  {title} {\enquote
  {\bibinfo {title} {Optimal measurements for simultaneous quantum estimation
  of multiple phases},}\ }\href@noop {} {\bibfield  {journal} {\bibinfo
  {journal} {Physical Review Letters}\ }\textbf {\bibinfo {volume} {119}},\
  \bibinfo {pages} {130504} (\bibinfo {year} {2017})}\BibitemShut {NoStop}%
\bibitem [{\citenamefont {Roccia}\ \emph {et~al.}(2017)\citenamefont {Roccia},
  \citenamefont {Gianani}, \citenamefont {Mancino}, \citenamefont {Sbroscia},
  \citenamefont {Somma}, \citenamefont {Genoni},\ and\ \citenamefont
  {Barbieri}}]{roccia2017entangling}%
  \BibitemOpen
  \bibfield  {author} {\bibinfo {author} {\bibfnamefont {E.}~\bibnamefont
  {Roccia}}, \bibinfo {author} {\bibfnamefont {I.}~\bibnamefont {Gianani}},
  \bibinfo {author} {\bibfnamefont {L.}~\bibnamefont {Mancino}}, \bibinfo
  {author} {\bibfnamefont {M.}~\bibnamefont {Sbroscia}}, \bibinfo {author}
  {\bibfnamefont {F.}~\bibnamefont {Somma}}, \bibinfo {author} {\bibfnamefont
  {M.~G.}\ \bibnamefont {Genoni}}, \ and\ \bibinfo {author} {\bibfnamefont
  {M.}~\bibnamefont {Barbieri}},\ }\bibfield  {title} {\enquote {\bibinfo
  {title} {Entangling measurements for multiparameter estimation with two
  qubits},}\ }\href@noop {} {\bibfield  {journal} {\bibinfo  {journal} {Quantum
  Science and Technology}\ }\textbf {\bibinfo {volume} {3}},\ \bibinfo {pages}
  {01LT01} (\bibinfo {year} {2017})}\BibitemShut {NoStop}%
\bibitem [{\citenamefont {Yousefjani}\ \emph {et~al.}(2017)\citenamefont
  {Yousefjani}, \citenamefont {Nichols}, \citenamefont {Salimi},\ and\
  \citenamefont {Adesso}}]{yousefjani2017estimating}%
  \BibitemOpen
  \bibfield  {author} {\bibinfo {author} {\bibfnamefont {R.}~\bibnamefont
  {Yousefjani}}, \bibinfo {author} {\bibfnamefont {R.}~\bibnamefont {Nichols}},
  \bibinfo {author} {\bibfnamefont {S.}~\bibnamefont {Salimi}}, \ and\ \bibinfo
  {author} {\bibfnamefont {G.}~\bibnamefont {Adesso}},\ }\bibfield  {title}
  {\enquote {\bibinfo {title} {Estimating phase with a random generator:
  Strategies and resources in multiparameter quantum metrology},}\ }\href@noop
  {} {\bibfield  {journal} {\bibinfo  {journal} {Physical Review A}\ }\textbf
  {\bibinfo {volume} {95}},\ \bibinfo {pages} {062307} (\bibinfo {year}
  {2017})}\BibitemShut {NoStop}%
\bibitem [{\citenamefont {Nichols}\ \emph {et~al.}(2018)\citenamefont
  {Nichols}, \citenamefont {Liuzzo-Scorpo}, \citenamefont {Knott},\ and\
  \citenamefont {Adesso}}]{nichols2017multiparameter}%
  \BibitemOpen
  \bibfield  {author} {\bibinfo {author} {\bibfnamefont {R.}~\bibnamefont
  {Nichols}}, \bibinfo {author} {\bibfnamefont {P.}~\bibnamefont
  {Liuzzo-Scorpo}}, \bibinfo {author} {\bibfnamefont {P.~A.}\ \bibnamefont
  {Knott}}, \ and\ \bibinfo {author} {\bibfnamefont {G.}~\bibnamefont
  {Adesso}},\ }\bibfield  {title} {\enquote {\bibinfo {title} {Multiparameter
  gaussian quantum metrology},}\ }\href@noop {} {\bibfield  {journal} {\bibinfo
   {journal} {Physical Review A}\ }\textbf {\bibinfo {volume} {98}},\ \bibinfo
  {pages} {012114} (\bibinfo {year} {2018})}\BibitemShut {NoStop}%
\bibitem [{\citenamefont {Proctor}\ \emph {et~al.}(2018)\citenamefont
  {Proctor}, \citenamefont {Knott},\ and\ \citenamefont
  {Dunningham}}]{proctor2018multiparameter}%
  \BibitemOpen
  \bibfield  {author} {\bibinfo {author} {\bibfnamefont {T.~J.}\ \bibnamefont
  {Proctor}}, \bibinfo {author} {\bibfnamefont {P.~A.}\ \bibnamefont {Knott}},
  \ and\ \bibinfo {author} {\bibfnamefont {J.~A.}\ \bibnamefont {Dunningham}},\
  }\bibfield  {title} {\enquote {\bibinfo {title} {Multiparameter estimation in
  networked quantum sensors},}\ }\href@noop {} {\bibfield  {journal} {\bibinfo
  {journal} {Physical Review Letters}\ }\textbf {\bibinfo {volume} {120}},\
  \bibinfo {pages} {080501} (\bibinfo {year} {2018})}\BibitemShut {NoStop}%
\bibitem [{\citenamefont {Nair}(2018)}]{nair2018quantum}%
  \BibitemOpen
  \bibfield  {author} {\bibinfo {author} {\bibfnamefont {R.}~\bibnamefont
  {Nair}},\ }\bibfield  {title} {\enquote {\bibinfo {title} {Quantum-limited
  loss sensing: Multiparameter estimation and bures distance between loss
  channels},}\ }\href@noop {} {\bibfield  {journal} {\bibinfo  {journal} {arXiv
  preprint arXiv:1804.02211}\ } (\bibinfo {year} {2018})}\BibitemShut {NoStop}%
\bibitem [{\citenamefont {Labeyrie}\ \emph {et~al.}(2006)\citenamefont
  {Labeyrie}, \citenamefont {Lipson},\ and\ \citenamefont
  {Nisenson}}]{labeyrie2006introduction}%
  \BibitemOpen
  \bibfield  {author} {\bibinfo {author} {\bibfnamefont {A.}~\bibnamefont
  {Labeyrie}}, \bibinfo {author} {\bibfnamefont {S.~G.}\ \bibnamefont
  {Lipson}}, \ and\ \bibinfo {author} {\bibfnamefont {P.}~\bibnamefont
  {Nisenson}},\ }\href@noop {} {\emph {\bibinfo {title} {An introduction to
  optical stellar interferometry}}}\ (\bibinfo  {publisher} {Cambridge
  University Press},\ \bibinfo {year} {2006})\BibitemShut {NoStop}%
\bibitem [{\citenamefont {Zmuidzinas}(2003)}]{zmuidzinas2003cramer}%
  \BibitemOpen
  \bibfield  {author} {\bibinfo {author} {\bibfnamefont {J.}~\bibnamefont
  {Zmuidzinas}},\ }\bibfield  {title} {\enquote {\bibinfo {title} {Cramer--rao
  sensitivity limits for astronomical instruments: implications for
  interferometer design},}\ }\href@noop {} {\bibfield  {journal} {\bibinfo
  {journal} {JOSA A}\ }\textbf {\bibinfo {volume} {20}},\ \bibinfo {pages}
  {218} (\bibinfo {year} {2003})}\BibitemShut {NoStop}%
\bibitem [{\citenamefont {Goodman}(2015)}]{goodman2015statistical}%
  \BibitemOpen
  \bibfield  {author} {\bibinfo {author} {\bibfnamefont {J.~W.}\ \bibnamefont
  {Goodman}},\ }\href@noop {} {\emph {\bibinfo {title} {Statistical optics}}}\
  (\bibinfo  {publisher} {John Wiley \& Sons},\ \bibinfo {year}
  {2015})\BibitemShut {NoStop}%
\bibitem [{\citenamefont {Mandel}\ and\ \citenamefont
  {Wolf}(1995)}]{mandel1995optical}%
  \BibitemOpen
  \bibfield  {author} {\bibinfo {author} {\bibfnamefont {L.}~\bibnamefont
  {Mandel}}\ and\ \bibinfo {author} {\bibfnamefont {E.}~\bibnamefont {Wolf}},\
  }\href@noop {} {\emph {\bibinfo {title} {Optical coherence and quantum
  optics}}}\ (\bibinfo  {publisher} {Cambridge University Press},\ \bibinfo
  {year} {1995})\BibitemShut {NoStop}%
\bibitem [{\citenamefont {Mandel}(1959)}]{mandel1959fluctuations}%
  \BibitemOpen
  \bibfield  {author} {\bibinfo {author} {\bibfnamefont {L.}~\bibnamefont
  {Mandel}},\ }\bibfield  {title} {\enquote {\bibinfo {title} {Fluctuations of
  photon beams: the distribution of the photo-electrons},}\ }\href@noop {}
  {\bibfield  {journal} {\bibinfo  {journal} {Proceedings of the Physical
  Society}\ }\textbf {\bibinfo {volume} {74}},\ \bibinfo {pages} {233}
  (\bibinfo {year} {1959})}\BibitemShut {NoStop}%
\bibitem [{\citenamefont {Gottesman}\ \emph {et~al.}(2012)\citenamefont
  {Gottesman}, \citenamefont {Jennewein},\ and\ \citenamefont
  {Croke}}]{gottesman2012longer}%
  \BibitemOpen
  \bibfield  {author} {\bibinfo {author} {\bibfnamefont {D.}~\bibnamefont
  {Gottesman}}, \bibinfo {author} {\bibfnamefont {T.}~\bibnamefont
  {Jennewein}}, \ and\ \bibinfo {author} {\bibfnamefont {S.}~\bibnamefont
  {Croke}},\ }\bibfield  {title} {\enquote {\bibinfo {title} {Longer-baseline
  telescopes using quantum repeaters},}\ }\href@noop {} {\bibfield  {journal}
  {\bibinfo  {journal} {Physical Review Letters}\ }\textbf {\bibinfo {volume}
  {109}},\ \bibinfo {pages} {070503} (\bibinfo {year} {2012})}\BibitemShut
  {NoStop}%
\bibitem [{\citenamefont {Tsang}(2011)}]{tsang2011quantum}%
  \BibitemOpen
  \bibfield  {author} {\bibinfo {author} {\bibfnamefont {M.}~\bibnamefont
  {Tsang}},\ }\bibfield  {title} {\enquote {\bibinfo {title} {Quantum
  nonlocality in weak-thermal-light interferometry},}\ }\href@noop {}
  {\bibfield  {journal} {\bibinfo  {journal} {Physical Review Letters}\
  }\textbf {\bibinfo {volume} {107}},\ \bibinfo {pages} {270402} (\bibinfo
  {year} {2011})}\BibitemShut {NoStop}%
\bibitem [{Note1()}]{Note1}%
  \BibitemOpen
  \bibinfo {note} {However, the assumption of weak sources can be lifted as in
  \cite {nair2016far,lupo2016ultimate}, leading to similar qualitative results
  for thermal sources of arbitrary intensity.}\BibitemShut {Stop}%
\bibitem [{\citenamefont {Tsang}(2015)}]{tsang2015quantum}%
  \BibitemOpen
  \bibfield  {author} {\bibinfo {author} {\bibfnamefont {M.}~\bibnamefont
  {Tsang}},\ }\bibfield  {title} {\enquote {\bibinfo {title} {Quantum limits to
  optical point-source localization},}\ }\href@noop {} {\bibfield  {journal}
  {\bibinfo  {journal} {Optica}\ }\textbf {\bibinfo {volume} {2}},\ \bibinfo
  {pages} {646} (\bibinfo {year} {2015})}\BibitemShut {NoStop}%
\bibitem [{\citenamefont {Svelto}\ and\ \citenamefont
  {Hanna}(1998)}]{svelto1998principles}%
  \BibitemOpen
  \bibfield  {author} {\bibinfo {author} {\bibfnamefont {O.}~\bibnamefont
  {Svelto}}\ and\ \bibinfo {author} {\bibfnamefont {D.~C.}\ \bibnamefont
  {Hanna}},\ }\href@noop {} {\emph {\bibinfo {title} {Principles of lasers}}}\
  (\bibinfo  {publisher} {Springer},\ \bibinfo {year} {1998})\BibitemShut
  {NoStop}%
\bibitem [{\citenamefont {Yang}\ \emph {et~al.}(2018)\citenamefont {Yang},
  \citenamefont {Pang}, \citenamefont {Zhou},\ and\ \citenamefont
  {Jordan}}]{yang2018optimal}%
  \BibitemOpen
  \bibfield  {author} {\bibinfo {author} {\bibfnamefont {J.}~\bibnamefont
  {Yang}}, \bibinfo {author} {\bibfnamefont {S.}~\bibnamefont {Pang}}, \bibinfo
  {author} {\bibfnamefont {Y.}~\bibnamefont {Zhou}}, \ and\ \bibinfo {author}
  {\bibfnamefont {A.~N.}\ \bibnamefont {Jordan}},\ }\bibfield  {title}
  {\enquote {\bibinfo {title} {Optimal measurements for quantum multi-parameter
  estimation with general states},}\ }\href@noop {} {\bibfield  {journal}
  {\bibinfo  {journal} {arXiv preprint arXiv:1806.07337}\ } (\bibinfo {year}
  {2018})}\BibitemShut {NoStop}%
\bibitem [{\citenamefont {Wolf}(2011)}]{wolf2011coherent}%
  \BibitemOpen
  \bibfield  {author} {\bibinfo {author} {\bibfnamefont {J.}~\bibnamefont
  {Wolf}},\ }\bibfield  {title} {\enquote {\bibinfo {title} {Coherent quantum
  control in biological systems},}\ }in\ \href@noop {} {\emph {\bibinfo
  {booktitle} {Biophotonics: Spectroscopy, Imaging, Sensing, and
  Manipulation}}}\ (\bibinfo  {publisher} {Springer},\ \bibinfo {year} {2011})\
  pp.\ \bibinfo {pages} {183--201}\BibitemShut {NoStop}%
\bibitem [{\citenamefont {Yu}\ and\ \citenamefont {Prasad}(2018)}]{YU1}%
  \BibitemOpen
  \bibfield  {author} {\bibinfo {author} {\bibfnamefont {Z.}~\bibnamefont
  {Yu}}\ and\ \bibinfo {author} {\bibfnamefont {S.}~\bibnamefont {Prasad}},\
  }\bibfield  {title} {\enquote {\bibinfo {title} {Quantum limited
  superresolution of an incoherent source pair in three dimensions},}\
  }\href@noop {} {\bibfield  {journal} {\bibinfo  {journal} {Physical Review
  Letters}\ }\textbf {\bibinfo {volume} {121}},\ \bibinfo {pages} {180504}
  (\bibinfo {year} {2018})}\BibitemShut {NoStop}%
\end{thebibliography}%

\clearpage
\appendix
	\renewcommand\theequation{S\arabic{equation}}
	\renewcommand\thepage{SM-\arabic{page}}
	\setcounter{equation}{0}
	\setcounter{page}{1}
	\section*{SUPPLEMENTAL MATERIAL}	
\section{A.~Expanding $\rho_1$ and  its derivatives in a non-orthogonal basis}
We start by observing that $\rho_1$ and all its derivatives --- and hence the associated SLDs --- are all supported in the subspace spanned by the vectors $\ket{\Psi_1}$, $\ket{\Psi_2}$ together with
\begin{align*}
\ket{\Psi_3}&\equiv\partial_{ x_1}\ket{\Psi_1}\\
\ket{\Psi_4}&\equiv\partial_{ z_1}\ket{\Psi_1}\\
\ket{\Psi_5}&\equiv\partial_{ x_2}\ket{\Psi_2}\\
\ket{\Psi_6}&\equiv\partial_{ z_2}\ket{\Psi_2}.
\end{align*}
We assume that the set $\{\ket{\Psi_j}\}_j$ is linearly independent provided that $x_1\neq x_2$ or $z_1\neq z_2$ (this may be always achieved up to appropriate limiting procedures), but the set is not orthonormal in general. Yet, such non-orthogonal basis can still be used to linearly expand any state or operator. The expressions to follow will all depend on the matrix $S$ of scalar products between the basis elements:
\begin{equation}
S_{ij}\equiv\sprod{\Psi_i}{\Psi_j}.
\end{equation}
Using the representation $\ket{\Psi_j}=\exp\left(-iG z_j-x_j\partial_x\right)\ket{\psi}$ provided in the main text --- and exploting that $G$ commutes with all $x-$ and $y-$ derivatives, we find that all the overlaps $S_{ij}$ depend only on the separations $s=x_2-x_1,p=z_2-z_1$, and not on the centroid coordinates. For example:
\begin{align}
\gamma&\equiv\sprod{\Psi_1}{\Psi_2}=\braket{\psi|e^{iG z_1+x_1\partial_x}e^{-iG z_2-x_2\partial_x}|\psi}\nonumber\\
&=\braket{\psi|e^{-iG p-s\partial_x}|\psi},
\end{align}
where we have also used that $\partial_x$ is anti-Hermitian. Similar simplifications, together with the paraxial wave equation $\partial_{ z_j}\ket{\Psi_j}=-iG\ket{\Psi_j}$, allow us to write all the matrix elements of $S$ as per
\begin{align}
S=\begin{pmatrix}
1&\gamma&0&-i\av{G}&\partial_{x_2}\gamma &\partial_{z_2}\gamma\\
\gamma^*&1&\partial_{ x_1}\gamma^*&\partial_{ z_1}\gamma^*&0&-i\av{G}\\
0&\partial_{x_1}\gamma &\braket{\partial_{x}\psi|\partial_{x}\psi}&0&\partial_{ x_1}\partial_{ x_2}\gamma&\partial_{ x_1}\partial_{ z_2}\gamma\\
i\av{G}&\partial_{z_1}\gamma &0&\av{G^2}&\partial_{z_1}\partial_{x_2}\gamma&\partial_{ z_1}\partial_{ z_2}\gamma\\
\partial_{ x_2}\gamma^*&0&\partial_{ x_1}\partial_{ x_2}\gamma^*&\partial_{z_1}\partial_{x_2}\gamma^*&\braket{\partial_{x}\psi|\partial_{x}\psi}&0\\
\partial_{ z_2}\gamma^*&i\av{G}&\partial_{ x_1}\partial_{ z_2}\gamma^*&\partial_{ z_1}\partial_{ z_2}\gamma^*&0&\av{G^2}
\end{pmatrix},
\end{align}
where it is worth noting that only $\gamma=\gamma(s,p)$ depends on the source separations, while $\braket{\partial_{x}\psi|\partial_{x}\psi}$, $\av{G}$ and $\av{G^2}$ are independent of all source parameters. In the above we exploited the further simplification $\braket{\psi|\partial_x\psi}=0,$ which follows from the assumption that $\psi(x,y,0)$ is real (see main text).

In the non-orthogonal basis $\{\ket{\Psi_j}\}_j$, the matrix representation of $\rho_1$ reads
\begin{equation}
\rho_1=\frac12\begin{pmatrix}
1&S_{12}&S_{13}&S_{14}&S_{15}&S_{16}\\
S_{21}&1&S_{23}&S_{24}&S_{25}&S_{26}\\
0&0&0&0&0&0\\
0&0&0&0&0&0\\
0&0&0&0&0&0\\
0&0&0&0&0&0
\end{pmatrix}.
\end{equation}
Since the above expression may at first sight appear strange, we emphasize that in a non-orthogonal basis, $\rho=\rho^\dagger$ \textit{does not imply} $\rho_{ij}=\rho_{ji}^*$. 
The derivatives of $\rho$ are instead represented by the matrices
\begin{align}
2\partial_{ x_1}\rho_1=\ketbra{\Psi_1}{\Psi_3}+\text{H.c.}=\begin{pmatrix}
S_{31}&S_{32}&S_{33}&S_{34}&S_{35}&S_{36}\\
0&0&0&0&0&0\\
1&S_{12}&S_{13}&S_{14}&S_{15}&S_{16}\\
0&0&0&0&0&0\\
0&0&0&0&0&0\\
0&0&0&0&0&0
\end{pmatrix}
\end{align}
\begin{align}
2\partial_{ z_1}\rho_1=\ketbra{\Psi_1}{\Psi_4}+\text{H.c.}=\begin{pmatrix}
S_{41}&S_{42}&S_{43}&S_{44}&S_{45}&S_{46}\\
0&0&0&0&0&0\\
0&0&0&0&0&0\\
1&S_{12}&S_{13}&S_{14}&S_{15}&S_{16}\\
0&0&0&0&0&0\\
0&0&0&0&0&0
\end{pmatrix}
\end{align}
\begin{align}
2\partial_{ x_2}\rho_1=\ketbra{\Psi_2}{\Psi_5}+\text{H.c.}=\begin{pmatrix}
0&0&0&0&0&0\\
S_{51}&S_{52}&S_{53}&S_{54}&S_{55}&S_{56}\\
0&0&0&0&0&0\\
0&0&0&0&0&0\\
S_{21}&1&S_{23}&S_{24}&S_{25}&S_{26}\\
0&0&0&0&0&0
\end{pmatrix}
\end{align}
\begin{align}
2\partial_{ z_2}\rho_1=\ketbra{\Psi_2}{\Psi_6}+\text{H.c.}=\begin{pmatrix}
0&0&0&0&0&0\\
S_{61}&S_{62}&S_{63}&S_{64}&S_{65}&S_{66}\\
0&0&0&0&0&0\\
0&0&0&0&0&0\\
0&0&0&0&0&0\\
S_{21}&1&S_{23}&S_{24}&S_{25}&S_{26}
\end{pmatrix}
\end{align}
We can now employ a symbolic manipulation software (e.g. Mathematica) to solve explicitly the SLD equations $L_\mu\rho+\rho L_\mu=2\partial_\mu\rho$, which gives us the matrix representation of the SLDs $(L_{x_1},L_{x_2},L_{z_1},L_{z_2})$ in the basis $\{\ket{\Psi_j}\}_j$. To find the SLDs corresponding to the variables of interest $(s,\bar x,p,\bar z)$ we then simply apply the rotation
\begin{equation}
\begin{pmatrix}
L_s\\L_{\bar x}\\L_p\\L_{\bar z}
\end{pmatrix}=\left(
\begin{array}{cccc}
-\frac{1}{2} & \frac{1}{2} & 0 & 0 \\
1 & 1 & 0 & 0 \\
0 & 0 & -\frac{1}{2} & \frac{1}{2} \\
0 & 0 & 1 & 1 \\
\end{array}
\right)\begin{pmatrix}
L_{x_1}\\L_{x_2}\\L_{z_1}\\L_{z_2}
\end{pmatrix}
\end{equation}
Once the SLDs are known, the results presented in the main text follow from the relation
\begin{equation}H_{\mu\nu}+i\Gamma_{\mu\nu}=\tr{[\rho_1L_\mu L_\nu]}\,.\end{equation}

\clearpage
\begin{widetext}
\section{B.~Explicit expressions for Gaussian beams}

Here we report the explicit expressions of the nonzero elements of the matrices $H_{\mu \nu}$ and $\Gamma_{\mu \nu}$ for Gaussian point spread functions specified by the parameters $k$ and $z_R$ as discussed in the main text. In the following we set $$\varsigma \equiv \frac{2 k s^2 z_R}{p^2+4 z_R^2}\,.$$

We have then

\begin{eqnarray}
H_{ss}&=&\frac{k}{2 z_R}\,, \nonumber\\
H_{{\bar x}{\bar x}} &=& \frac{\varsigma}{s^2}  \left(p^2 \left(\frac{1}{z_R^2}-\frac{2
   \varsigma ^2}{k e^{\varsigma } s^2 z_R-2 \varsigma
   z_R^2}\right)-\frac{8 e^{-\varsigma } \varsigma ^2
   z_R}{k s^2}+4\right)\,, \nonumber\\
H_{pp} &=& \frac{1}{4z_R^2}\,, \nonumber\\
   H_{{\bar z}{\bar z}} &=& \frac{e^{-\varsigma } \left(4 k^4 e^{2 \varsigma } s^8-k^2
   e^{\varsigma } \varsigma ^2 s^4 \left(p^2 (\varsigma^2
   -4 \varsigma +8)+4 (\varsigma ^2+4)
   z_R^2\right)+16 p^2 (\varsigma -1)^2 \varsigma ^4
   z_R^2\right)}{4 k^3 s^6 z_R^2 \left(k e^{\varsigma }
   s^2-2 \varsigma  z_R\right)}\,, \nonumber\\
   H_{{\bar x}{\bar z}} &=& \frac{p e^{-\varsigma } \varsigma ^2 \left(k^2 e^{\varsigma
   } (\varsigma -2) s^4-8 (\varsigma -1) \varsigma ^2
   z_R^2\right)}{k^2 s^5 z_R \left(k e^{\varsigma } s^2-2
   \varsigma  z_R\right)}\,,
\end{eqnarray}
and
\begin{eqnarray}
\Gamma_{s{\bar x}} &=& -\frac{4 p e^{-\varsigma } \varsigma ^4 z_R}{k^2
   e^{\varsigma } s^6-2 k \varsigma  s^4 z_R} \,,\nonumber\\
\Gamma_{p{\bar z}} &=& -\frac{p e^{-\varsigma } (\varsigma -1) \varsigma ^4
   \left(p^2 (\varsigma -2)-4 \varsigma  z_R^2\right)}{2
   k^3 s^6 z_R \left(k e^{\varsigma } s^2-2 \varsigma
   z_R\right)}\,,\nonumber\\
\Gamma_{s{\bar z}} &=& \frac{e^{-\varsigma } \varsigma ^3 \left(2 p^2 (\varsigma
   -1) \varsigma -k^2 e^{\varsigma } s^4\right)}{k^2 s^5
   \left(k e^{\varsigma } s^2-2 \varsigma  z_R\right)}\,,\nonumber\\
\Gamma_{{\bar x}p} &=& -\frac{e^{-\varsigma } \varsigma ^3 \left(k^2 e^{\varsigma
   } s^4+\varsigma  \left(p^2 (\varsigma -2)-4 \varsigma
   z_R^2\right)\right)}{k^2 s^5 \left(k e^{\varsigma }
   s^2-2 \varsigma  z_R\right)} \,.
\end{eqnarray}

\vfill
\clearpage	
\end{widetext}

\end{document}